\documentclass[12pt,draftclsnofoot, onecolumn]{IEEEtran}
\usepackage{graphicx,cite,epsfig}
\usepackage{amsmath,amssymb}
\usepackage{amsmath,bm}
\usepackage{setspace}

\begin{document}
\markboth{IEEE Journal on Selected Areas in Communications, Vol. XX,
No. Y, Month 2016} {Ge etc.: Millimeter Wave Communications with OAM-SM Scheme for Future Mobile Networks\ldots}
\title{\mbox{}\vspace{0.40cm}\\
\textsc{Millimeter Wave Communications with OAM-SM Scheme for Future Mobile Networks} \vspace{0.2cm}}

\author{\normalsize
Xiaohu Ge$^1$, Ran Zi$^1$, Xusheng Xiong$^1$, Qiang Li$^1$, Liang Wang$^1$\\
\vspace{0.70cm} \small{
$^1$School of Electronic Information and Communications\\
Huazhong University of Science and Technology, Wuhan 430074, Hubei, P. R.
China.\\}
Email: \{xhge, ranzi\_pds, xiongxusheng, qli\_patrick, M201671825\}@mail.hust.edu.cn\\
\vspace{0.1cm}


\thanks{\small{Submitted to IEEE JSAC Special Issue on Millimeter Wave Communications for Future Mobile Networks.}}
\thanks{\small{Correspondence author: Dr. Qiang Li, Tel: +86 (0)27 87557942, Fax: +86 (0)27 87557943, Email: qli\_patrick@mail.hust.edu.cn.}}
\thanks{\small{The authors would like to acknowledge the support from the National Natural Science Foundation of China (NSFC) under the grants 61271224, NFSC Major International Joint Research Project under the grant 61210002, the Fundamental Research Funds for the Central Universities under the grant 2015XJGH011. This research is partially supported by the EU FP7-PEOPLE-IRSES, project acronym CROWN (grant no. 610524), project acronym WiNDOW (grant no. 318992), China International Joint Research Center of Green Communications and Networking (No. 2015B01008).}}
}

\date{\today}
\renewcommand{\baselinestretch}{1.2}
\thispagestyle{empty} \maketitle \thispagestyle{empty}
\newpage
\setcounter{page}{1}\begin{abstract}

The orbital angular momentum (OAM) technique provides a new degree of freedom for information transmissions in millimeter wave communications. Considering the spatial distribution characteristics of OAM beams, a new OAM spatial modulation (OAM-SM) millimeter wave communication system is first proposed for future mobile networks. Furthermore, the capacity, average bit error probability and energy efficiency of OAM-SM millimeter wave communication systems are analytically derived for performance analysis. Compared with the conventional multi-input multi-output (MIMO) millimeter wave communication systems, the maximum capacity and energy efficiency of OAM-SM millimeter wave communication systems are improved by 36\% and 472.3\%, respectively. Moreover, numerical results indicate that the proposed OAM-SM millimeter wave communication systems are more robust to path-loss attenuations than the conventional MIMO millimeter wave communication systems, which makes it suitable for long-range transmissions. Therefore, OAM-SM millimeter wave communication systems provide a great growth space for future mobile networks.

\end{abstract}
\newpage
\IEEEpeerreviewmaketitle \vspace{-1cm}

\begin{spacing}{1.405}
\section{Introduction}
\label{sec1}

To meet the 1000 times growth of wireless traffic in the next decade, the millimeter wave transmission technology is becoming one of the key technologies for the future fifth generation (5G) mobile networks \cite{Thompson14}. With the implementation of millimeter wave transmission technologies, a lot of frequency bandwidths are utilized for 5G mobile communication systems \cite{Rappaport13}. Moreover, the high spectrums, {\em e.g.}, 100-300 GHz spectrums are proposed for future mobile networks \cite{Ge15,Akdeniz14}. Consequently, the orbital angular momentum (OAM) technology traditionally used in optical communications is explored to be used for millimeter wave communications in the high spectrums \cite{Yan14}. Compared with the conventional degree of freedoms in wireless communications, such as degree of freedoms in time, frequency and space, the OAM technology provides a new degree of freedom, {\em i.e.}, the OAM state for millimeter wave communications \cite{Yao13}. A revolutionary improvement on millimeter wave communications can be expected for future mobile networks when the OAM technology is adopted for millimeter wave communications.

Based on the electromagnetic theory, an electromagnetic wave not only has the linear momentum along the beam propagation direction but also has the angular momentum which is perpendicular to the beam propagation direction \cite{Jackson98}. The OAM is one part of the angular momentum whose electromagnetic fields are characterized by a wavefront that is shaped as a helix, {\em i.e.}, $exp \left( { - il\phi } \right)$, where $\phi $ is the transverse azimuthal angle and $l$ is the OAM state, {\em i.e.}, an unbounded integer. The OAM state $l$ implies a $2\pi l$ phase delay over one revolution of a beam in millimeter wave communications. Moreover, beams with different OAM states are mutually orthogonal and thus can be multiplexed together along the same beam axis for propagation. Thereby, the capacity of wireless communication systems could be obviously improved by the OAM technology \cite{Jackson98,Gibson04}. Although the OAM feature of electromagnetic wave has been indicated in early years of $20^{th}$ century \cite{Beth36}, until recent years the OAM technology has been used for optical communications \cite{Gibson04,Allen92,Awaji10,Djordjevic11,Wang11,Fazal11,Bozinovic13}. Laser light with Laguerre-Gaussian amplitude distribution was first found to have an OAM and the Laguerre-Gaussian mode can be transformed from the Hermite-Gaussian mode in astigmatic optical systems \cite{Allen92}. Utilizing the OAM technology, information was encoded into a light beam and was transmitted by free space communication experiments \cite{Gibson04}. Using Laguerre-Gaussian mode beam, the spatial mode division multiplexing with two OAM states was first demonstrated to achieve 10 Gbps transmission rate in optical communications \cite{Awaji10}. By Combining with low-density parity-check (LDPC) codes, OAM-based modulation schemes were proposed to realize high speed transmissions over strong atmospheric turbulence channels in deep-space and near-earth optical communications \cite{Djordjevic11}. Modulating 16-quadrature-amplitude-modulation signals over four OAM states, a spectral efficiency of 25.6 bit/s/Hz was implemented by polarization-multiplexed multiple OAM optical communication systems \cite{Wang11}. A 2-Tbit/s data link was presented in free space optical communication systems using two orthogonal OAM states with 25 wavelength division multiplexed channels on each state \cite{Fazal11}. In a 1.1 kilometer transmission scenario over an optical fiber, 1.6 Tbit/s data transmission rate was demonstrated for optical communications with two OAM states over 10 wavelengths \cite{Bozinovic13}.

In addition to optical communications, the OAM technology has also been implemented for wireless transmissions over millimeter wave and microwave spectrums. Thide {\em et al.}, first proposed that vector antenna arrays can generate radio beams with OAM characteristics which is similar with helical Laguerre-Gaussian laser beams in paraxial optics \cite{Thid¨¦07}. Thereby, the OAM technology was suggested to be used for wireless communications with microwave spectrums. Fabrizio {\em et al.}, first demonstrated that two beams of incoherent radio waves, transmitted on the same frequency but encoded in two different OAM states, simultaneously transmit two independent radio channels \cite{Tamburini12}. Experiments results in \cite{Tamburini12} have enkindled discussions about the comparison between the OAM and multi-input multi-output (MIMO) systems \cite{Edfors12,Zhang13,Andersson15}. Based on the configuration of uniform circular antenna (UCA) array in free space communications, Ove {\em et al.}, indicated that communicating over the sub-channels given by OAM states is a subset of solutions offered by MIMO technology \cite{Edfors12}. When the transmitting and receiving antenna arrays have not been aligned in free space, the capacity of OAM-based radio communication decreases significantly with disalignment antenna arrays and the rotation phase is seriously distorted \cite{Zhang13}. Moreover, the impact of crosstalk and poor signal-to-noise ratio (SNR) on OAM-base radio communications has been discussed in \cite{Andersson15}. The above studies of OAM radio communications are based on UCA arrays. To overcome drawbacks of UCA arrays, Yan {\em et al.}, utilized the spiral phase plates (SPP) to generate OAM beams and implemented a transmission rate of 32 Gbps, {\em i.e.}, 16 bit/s/Hz spectral efficiency, over millimeter wave channels multiplexing eight OAM states \cite{Yan14}. However, it is very complex to generate OAM beams by SPP systems. To solve this problem, a dual OAM antenna with traveling-wave ring resonators was designed to simultaneously generate multiple beams with different OAM states for millimeter wave communications \cite{Hui15}. Furthermore, the technology of using uniform linear array, which consists of circular traveling-wave OAM antennas for multiplexing, was compared with the conventional MIMO communication method, and simulation results implied that OAM waves could decrease the spatial correlation of MIMO channels \cite{Zhang16}. Based on the property that the intensity of electric field in radial direction is characterized by the Bessel function of the first kind, three methods were proposed for beam axis detection and alignment for OAM-based radio communications \cite{Tian16}. Utilizing OAM states, a basic method was proposed for encoding and decoding data in wireless communications \cite{Allen14}. However, most of the above studies on OAM-based radio communications are focused on multiplexing OAM orthogonal channels. When the OAM technology is limited in the channel multiplexing, the capacity of OAM-based radio communication systems can not be larger than the capacity of MIMO radio communication systems \cite{Oldoni15}. Although the method in \cite{Allen14} provided a potential approach to extend a new degree of freedom in wireless communication, the total communication system has not been explained in detail and the corresponding performance analysis is insufficient.

The assumption that gains of OAM beams are the same in different propagation directions is widely adopted in current OAM-based radio communications. But this assumption conflicts with the spatial distribution characteristic of OAM beams where the energy of OAM beam is focused on a circle surrounding the beam axis \cite{Yao13,Allen92}. Considering the spatial distribution characteristic of OAM beams, in this paper a new OAM spatial modulation (OAM-SM) scheme is proposed for multi-antenna millimeter wave communication systems. Main contributions of this paper are list as follows:
\begin{enumerate}
\item A new OAM-SM scheme is proposed for multi-antenna millimeter wave communication systems considering the spatial distribution characteristic of OAM beams.
\item The capacity, average bit error probability (ABEP) and energy efficiency of OAM-SM millimeter wave communication systems are analytically derived in closed forms. Moreover, the capacity, ABEP and energy efficiency of OAM-SM millimeter wave communication systems are simulated for performance analysis.
\item Compared with the conventional multi-input multi-output (MIMO) millimeter wave communication systems, the maximum capacity and energy efficiency of OAM-SM millimeter wave communication systems are improved by 36\% and 472.3\%, respectively. Moreover, numerical results indicate that the capacity of OAM-SM millimeter wave communication systems is larger than the capacity of MIMO millimeter wave communication systems when the transmission distance is larger than 38 meters.

\end{enumerate}

The remainder of this paper is outlined as follows. Section II describes the system model of OAM-SM millimeter wave communications and the OAM wireless channel model is derived. The capacity of OAM-SM millimeter wave communication systems is derived and analyzed in Section III. The ABEP of OAM-SM millimeter wave communication systems is proposed and analyzed in Section IV. The energy efficiency of OAM-SM millimeter wave communication systems is presented and analyzed in Section V. Finally, conclusions are drawn in Section VI.

\section{System Model of OAM-SM Millimeter Wave Communications}
\label{sec2}
Based on the spatial distribution characteristic of OAM beams, an OAM-SM scheme is proposed for multi-antenna millimeter wave communication systems. In the following, we illustrate the system model of OAM-SM millimeter wave communications, including the OAM-SM scheme, the OAM wireless channel model and the OAM spatial demodulation scheme.

\subsection{OAM-SM Scheme}

The system model of OAM-SM millimeter wave communications is illustrated in Fig. 1, where $M$ transmission antennas are formed a uniform linear array for transmitting OAM beams. Without loss of generality, in this paper OAM antennas are configured with traveling-wave ring resonators to generate beams with different OAM states. To easy generate OAM signals, a millimeter wave spectrum is used for wireless transmission by the OAM-SM scheme. Considering the direction of OAM signals, the receive antennas are assumed to be located at circle regions of OAM signals. Moreover, receive antenna pairs are configured at the circle region of OAM signals. Every receive antenna pair, including two receive antennas, can detect the phase of OAM signals and then determine the OAM state of signals. Hence, every transmission antenna corresponds to a receive antenna pair in the proposed OAM-SM millimeter wave communication systems. In this case, there are $M$ pairs of receive antennas at the receiver of OAM-SM millimeter wave communication systems when the transmitter is equipped with as $M$ transmission antennas.

Based on the OAM-SM scheme, only one transmission antenna is adopted to transmit OAM beams in one time slot. In this case, ${\log _2}M$ bits information can be transmitted by selecting one antenna from $M$ transmission antennas in the OAM-SM scheme. When the number of OAM states is configured as $L$ for every OAM signal transmitted by a transmission antenna, ${\log _2}L$ bits information can be transmitted by selecting one state from $L$ OAM states. Moreover, an OAM signal is modulated by the conventional P-point constellation modulation, {\em i.e.}, the radiated symbol ${x_p} \in \mathcal{X}$, $1 \leqslant p \leqslant P$, where $\mathcal{X}$ is the set of radiated symbols corresponding the P-point constellation modulation. The value of ${x_p}$ is assigned with equivalent probability from $P$ candidate values in $\mathcal{X}$. Consequently, a symbol modulated at the OAM signal brings ${\log _2}P$ bits information and the power of the radiated symbol is assumed as $\mathbb{E}\left( {{{\left| {{x_p}} \right|}^2}} \right) = 1$. Therefore, ${\log _2}M + {\log _2}L + {\log _2}P$ bits information can be transmitted by an OAM signal in OAM-SM millimeter wave communication systems.
The receive signal of OAM-SM millimeter wave communication systems is expressed by
\[{\bf{y}} = \sqrt \rho  {\bf{h}}_m^l{x_p} + {\bf{w}},\tag{1}\]
where ${\mathbf{y}} = {\left[
{{{\mathbf{y}}_1},...{{\mathbf{y}}_m},...,{{\mathbf{y}}_j},...,{{\mathbf{y}}_M}} \right]^T} \in {\mathbb{C}^{M \times 2}}$, $1 \leqslant m \leqslant M$, $1 \leqslant j \leqslant M$, $j \ne m$, is the signal vector at $M$ pairs of receive antennas, ${{\mathbf{y}}_m} = \left[ {{y_{m,1}},{y_{m,2}}} \right]$ is the signal vector at the $m - th$ receive antenna pair, ${y_{m,1}}$ and ${y_{m,2}}$ correspond to the signals at the first and second antennas of the $m - th$ receive antenna pair  respectively, ${y_{j,1}}$ and ${y_{j,2}}$ correspond to the signals at the first and second antennas of the $j - th$ receive antenna pair respectively. $\rho$ is the transmission power of OAM-SM millimeter wave communication systems. ${\mathbf{h}}_m^l \in {\mathbb{C}^{M \times 2}}$ is the channel response between the receive antennas and the $m - th$ transmission antenna with the OAM state $l$, $l \in \mathcal{L}$ and $\mathcal{L}$ is the OAM state set. The channel response ${\mathbf{h}}_m^l$ is extended by
\[{\mathbf{h}}_m^l = {\left[ {{\mathbf{h}}_{m1}^l,...,{\mathbf{h}}_{mm}^l,...,{\mathbf{h}}_{mj}^l,...,{\mathbf{h}}_{mM}^l} \right]^T},\tag{2}\]
where ${\mathbf{h}}_{mm}^l = \left[ {h_{mm,1}^l,h_{mm,2}^l} \right]$ is the channel response between the $m - th$ transmission antenna and the $m - th$ receive antenna pair, $h_{mm,1}^l$ and $h_{mm,2}^l$ correspond to the channel responses at the first and second antennas of the $m - th$ receive antenna pair. ${\mathbf{h}}_{mj}^l = \left[ {h_{mj,1}^l,h_{mj,2}^l} \right]$ is the channel response between the $m - th$ transmission antenna and the $j - th$ receive antenna pair, $h_{mj,1}^l$ and $h_{mj,2}^l$ correspond the channel responses at the first and second antennas of the $j - th$ receive antenna pair. ${\mathbf{w}} = {\left[ {{{\mathbf{w}}_1},...{{\mathbf{w}}_m},...,{{\mathbf{w}}_j},...,{{\mathbf{w}}_M}} \right]^T}$ is the noise vector at OAM-SM wireless channels, ${{\mathbf{w}}_m} = \left[ {{w_{m,1}},{w_{m,2}}} \right]$ and ${{\mathbf{w}}_j} = \left[ {{w_{j,1}},{w_{j,2}}} \right]$ are the additional Gaussian white noise (AWGN) corresponding to channels at the $m - th$ and the $j - th$ receive antenna pairs. The AWGN is assumed to be governed by a random distribution with the mean zero and the variance $\sigma _{\mathbf{w}}^2$ in this paper. Based on the OAM-SM scheme, the OAM-SM wireless channel model is further derived as follows.

\subsection{OAM-SM Wireless Channel Models}

In this paper the position of antennas in Fig. 1 is expressed by the cylindrical coordinate system $\left( {r,\phi ,\mathfrak{Z}} \right)$, where $r$ is the polar axis, $\phi$ is the angular coordinate and $\mathfrak{Z}$ is the longitudinal axis. Based on the system model of OAM-SM millimeter wave communications in Fig. 1, ${\text{T}}{{\text{x}}_m}$, $1 \leqslant m \leqslant M$, is the $m - th$ transmission antenna, ${\text{R}}{{\text{x}}_{m,1}}$ and ${\text{R}}{{\text{x}}_{m,2}}$ are the first and second antennas of the $m - th$ receive antenna pair. The angle between the receive antenna pair is assumed as $\beta$ which corresponds to the azimuth in cylindrical coordinate systems. Based on results in \cite{Mohammadi10}, the accurate measurement of OAM states needs to satisfy a constraint $\beta  < \frac{\pi }{{\left| {{L_{\max }}} \right|}}$, where $\left| {{L_{\max }}} \right|$ is the maximum absolute value of the OAM state. Without loss of generality, the position of the ${\text{T}}{{\text{x}}_m}$ antenna is configured as the origin of cylindrical coordinate system in this paper. Correspondingly, the position of the ${\text{T}}{{\text{x}}_j}$ transmission antenna is expressed by $\left( {(j - m)\xi ,0,0} \right)$, where $\xi$ is the distance between adjacent antennas. Considering the spatial distribution characteristic of OAM beams, the beam axis of OAM signals is denoted by $\mathfrak{Z}$ axis in this paper. When parabolic reflectors are used for shaping OAM signals in OAM-SM millimeter wave communication systems \cite{Hui15,Zhang16}, OAM beams with different OAM states are assumed to have the same size of OAM circle regions. Furthermore, the positions of the $m - th$ receive antenna pair ${\text{R}}{{\text{x}}_{m,1}}$ and ${\text{R}}{{\text{x}}_{m,2}}$ are denoted by $\left( {{r_{\max }}\left( z \right),\frac{\pi }{2},z} \right)$ and $\left( {{r_{\max }}\left( z \right),\frac{\pi }{2}{\text{ + }}\beta ,z} \right)$, where ${r_{\max }}(z)$ is the radius of the OAM circle region when the OAM propagation distance is $z$ in the $\mathfrak{Z}$ axis. For the $j - th$ receive antenna pair, the distances between the antenna ${\text{T}}{{\text{x}}_m}$ and the receive antenna pair ${\text{R}}{{\text{x}}_{j,1}}$, ${\text{R}}{{\text{x}}_{j,2}}$ are denoted by
\[{d_{mj,1}} = \sqrt {{z^2} + {{(j - m)}^2}{\xi ^2} + r_{\max }^2(z)},\tag{3a}\]
\[{d_{mj,2}} = \sqrt {{z^2} + {{\left( {(j - m)\xi  - {r_{\max }}\left( z \right)\sin \beta } \right)}^2} + r_{\max }^2(z){{\cos }^2}\beta }.\tag{3b}\]

The azimuths of the $j - th$ receive antenna pair are denoted by
\[{\phi _{mj,1}} = \left\{ {\begin{array}{*{20}{c}}
  {\arctan \frac{{{r_{\max }}\left( z \right)}}{{\left| {j - m} \right|\xi }},j > m} \\
  \begin{gathered}
  \pi  - \arctan \frac{{{r_{\max }}\left( z \right)}}{{\left| {j - m} \right|\xi }},j < m \hfill \\
  \frac{\pi }{2},j = m \hfill \\
\end{gathered}
\end{array}} \right.,\tag{4a}\]
\[{\phi _{mj,2}} = \left\{ {\begin{array}{*{20}{c}}
  {\arctan \frac{{{r_{\max }}\left( z \right)\cos \beta }}{{\left| {j - m} \right|\xi  - {r_{\max }}\left( z \right)\sin \beta }},j > m} \\
  \begin{gathered}
  \pi  - \arctan \frac{{{r_{\max }}\left( z \right)\cos \beta }}{{\left| {j - m} \right|\xi  + {r_{\max }}\left( z \right)\sin \beta }},j < m \hfill \\
  \frac{\pi }{2} + \beta ,j = m \hfill \\
\end{gathered}
\end{array}} \right. .\tag{4b}\]

As a consequence, the positions of $j - th$ receive antenna pair ${\text{R}}{{\text{x}}_{j,1}}$ and ${\text{R}}{{\text{x}}_{j,2}}$ are expressed by $(\sqrt {{{(j - m)}^2}{\xi ^2} + r_{\max }^2(z)} ,{\phi _{mj,1}},z)$ and $(\sqrt {{{\left( {(j - m)\xi  - {r_{\max }}\left( z \right)\sin \beta } \right)}^2} + r_{\max }^2(z){{\cos }^2}\beta } ,{\phi _{mj,1}},z)$, respectively.

In the OAM circle regions, the fading of OAM signals is governed by the Friis Law \cite{Oldoni15}, {\em i.e.}, the strength of signal is attenuated with the inverse square of the propagation distance. Consequently, the channel responses between the transmission antenna ${\text{T}}{{\text{x}}_m}$ and the receive antenna pair ${\text{R}}{{\text{x}}_{m,1}}$, ${\text{R}}{{\text{x}}_{m,2}}$ are expressed by
\[h_{mm,1}^l = \mathcal{B}\frac{\lambda }{{4\pi {d_{mm,1}}}}{e^{ - ik{d_{mm,1}}}}{e^{ - i\frac{\pi }{4}l}},\tag{5a}\]
\[h_{mm,2}^l = \mathcal{B}\frac{\lambda }{{4\pi {d_{mm,2}}}}{e^{ - ik{d_{mm,2}}}}{e^{ - i\left( {\frac{\pi }{4} + \beta } \right)l}},\tag{5b}\]
where $\lambda$ is the wave length, $k$ is the wave number, {\em i.e.}, $k = \frac{{2\pi }}{\lambda }$, $i$ is the imaginary unit, ${e^{ - ik{d_{mn,1}}}}$ and ${e^{ - ik{d_{mn,2}}}}$ are the phase changes corresponding the receive antenna pair ${\text{R}}{{\text{x}}_{m,1}}$ and ${\text{R}}{{\text{x}}_{m,2}}$, ${e^{ - i\frac{\pi }{4}l}}$ and ${e^{ - i\left( {\frac{\pi }{4} + \beta } \right)l}}$ are the OAM helical phases corresponding the receive antenna pair ${\text{R}}{{\text{x}}_{m,1}}$ and ${\text{R}}{{\text{x}}_{m,2}}$, $\mathcal{B}$ is the gain coefficient of transmission antennas in circle regions which is configured as a constant considering the Friis law in OAM circle regions. The channel responses between the transmission antenna ${\text{T}}{{\text{x}}_m}$ and the receive antenna pair ${\text{R}}{{\text{x}}_{j,1}}$, ${\text{R}}{{\text{x}}_{j,2}}$, are expressed by
\[h_{mj,1}^l = {\mathcal{B}_{mj,1}}\frac{\lambda }{{4\pi {d_{mj,1}}}}{e^{ - ik{d_{mj,1}}}}{e^{ - il{\phi _{mj,1}}}},\tag{6a}\]
\[h_{mj,2}^l = {\mathcal{B}_{mj,2}}\frac{\lambda }{{4\pi {d_{mj,2}}}}{e^{ - ik{d_{mj,2}}}}{e^{ - il{\phi _{mj,2}}}},\tag{6b}\]
where ${\mathcal{B}_{mj,1}}$ and ${\mathcal{B}_{mj,2}}$ are the gain coefficients of the transmission antenna outside circle regions which correspond to the receive antenna pair ${\text{R}}{{\text{x}}_{j,1}}$ and ${\text{R}}{{\text{x}}_{j,2}}$. Since the receive antenna pair ${\text{R}}{{\text{x}}_{j,1}}$ and ${\text{R}}{{\text{x}}_{j,2}}$ are not located at the OAM circle region transmitted by the antenna ${\text{T}}{{\text{x}}_m}$, ${\mathcal{B}_{mj,1}}$ and ${\mathcal{B}_{mj,2}}$ are not a constant and derived in the following.

The strength distribution of OAM signals is described by Laguerre-Gaussian beam in the free space \cite{Yao13}. In cylindrical coordinate systems, the Laguerre-Gaussian beam is expressed by \cite{Allen92}
\[\begin{gathered}
  u(r,\phi ,z) = \gamma \sqrt {\frac{{\mathfrak{L}!}}{{\pi \left( {\mathfrak{L} + \left| l \right|} \right)!}}} \frac{1}{{{w_l}(z)}}{(\frac{{r\sqrt 2 }}{{{w_l}(z)}})^{\left| l \right|}}{e^{ - {{(\frac{r}{{{w_l}(z)}})}^2}}}\mathbb{L}_p^{\left| l \right|}(\frac{{2{r^2}}}{{w_l^2(z)}}) \\
   \cdot {e^{ - i\frac{{\pi {r^2}}}{{\lambda {R_l}(z)}}}}{e^{i(\left| l \right| + 2\mathfrak{L} + 1)\psi (z)}}{e^{ - il\phi }} \\
\end{gathered} ,\tag{7a}\]
\[{R_l}(z) = z\left[ {1 + {{(\frac{{\pi w_l^2}}{{\lambda z}})}^2}} \right],\tag{7b}\]
where $\gamma \sqrt {\frac{{\mathfrak{L}!}}{{\pi \left( {\mathfrak{L} + \left| l \right|} \right)!}}} $ is a normalized constant, $\mathfrak{L}$ is the radial index which describes the number of radial nodes in the intensity distribution, for the proposed OAM-SM millimeter wave communication systems, the radial index is configured as $\mathfrak{L} = 0$. $l$ is the OAM state, ${w_l}\left( z \right)$ is the radius at which the field amplitudes fall into 1/e of their axial values. The expression of ${w_l}\left( z \right)$ is extended by
\[{w_l}\left( z \right) = {w_l}\sqrt {1 + {{\left( {\frac{z}{{{z_R}}}} \right)}^2}},\tag{8}\]
where ${w_l}$ is the beam waist radius with the OAM state $l$ and $z = 0$. ${z_R} = \frac{{\pi w_l^2}}{\lambda }$ is the Rayleigh distance. $\mathbb{L}_p^{\left| l \right|}\left( {\frac{{2{r^2}}}{{{w^2}\left( z \right)}}} \right)$ is the generalized Laguerre polynomial, ${e^{ - il\phi }}$ is the helical phase distribution when the OAM state $l \ne 0$, $\psi (z) = arctan(\frac{z}{{{z_R}}})$ is the Gouy phase at $z$, {\em i.e.}, an extra phase term beyond that attributable to the phase velocity of signal. The energy of OAM signal is focused in the OAM circle region. The radius of the OAM circle region with the maximum energy strength is configured by
\[{r_{\max }}(z) = \sqrt {\frac{{\left| l \right|}}{2}} {w_l}(z) = {w_l}\sqrt {\frac{{\left| l \right|}}{2}\left( {1 + {{\left( {\frac{z}{{{z_R}}}} \right)}^2}} \right)} .\tag{9}\]

In this paper OAM beams with different OAM states are assumed to have the same size of OAM circle regions, {\em i.e.}, the same radius of OAM circle regions. Moreover, the ${w_l}$ with the OAM state $l$ is assumed as a constant. For other OAM signals with different OAM state $l'$, $l' \ne l$, $l' \in \mathcal{L}$, the following equation should be satisfied by
\[{w_l}\sqrt {\frac{{\left| l \right|}}{2}\left( {1 + {{\left( {\frac{z}{{{z_R}}}} \right)}^2}} \right)}  = {w_{l'}}\sqrt {\frac{{\left| {l'} \right|}}{2}\left( {1 + {{\left( {\frac{z}{{{z_R}}}} \right)}^2}} \right)}.\tag{10}\]
Substitute the Rayleigh distance ${z_R} = \frac{{\pi w_l^2}}{\lambda }$ into (10). Furthermore, (10) is transformed by
\[Aw_{l'}^4 - Bw_{l'}^2 + C = 0,\tag{11a}\]
with
\[A = w_l^2\left| l \right|{\pi ^2},\tag{11b}\]
\[B = \left| l \right|({\pi ^2}w_l^4 + {z^2}{\lambda ^2}),\tag{11c}\]
\[C = w_l^2\left| {l'} \right|{z^2}{\lambda ^2},\tag{11d}\]
Based on the solution of (11), the beam waist radius ${w_{l'}}$ corresponding the OAM state $l'$ is obtained. When the result of ${w_{l'}}$ is substituted into (7), the strength distribution of OAM signal with the OAM state $l'$ is derived. In this case, the $u(r,\phi ,z)$ is regarded as the response of OAM electromagnetic wave in the cylindrical coordinate system after a unit pulse is input. When the OAM beam is transmitted by the antenna ${\text{T}}{{\text{x}}_m}$ in OAM-SM millimeter wave communication systems, the response at the antenna ${\text{R}}{{\text{x}}_{m,1}}$ is denoted by ${u_{mm,1}} = h_{mm,1}^l\overset{\lower0.5em\hbox{$\smash{\scriptscriptstyle\frown}$}}{x}$ and the response at the antenna ${\text{R}}{{\text{x}}_{j,1}}$ is denoted by ${u_{mj,1}} = h_{mj,1}^l\overset{\lower0.5em\hbox{$\smash{\scriptscriptstyle\frown}$}}{x}$, where $\overset{\lower0.5em\hbox{$\smash{\scriptscriptstyle\frown}$}}{x} $ is the unit pulse input. Furthermore, the following equation is satisfied by
\[\frac{{{u_{mm,1}}}}{{{u_{mj,1}}}} = \frac{{h_{mm,1}^l}}{{h_{mj,1}^l}}.\tag{12}\]
The results of ${u_{mm,1}}$ and ${u_{mj,1}}$ can be derived by (7). Substitute the results of ${u_{mm,1}}$ and ${u_{mj,1}}$, (5), (6) into (12), the gain coefficients of the transmission antenna corresponding the receive antenna ${\text{R}}{{\text{x}}_{j,1}}$ is derived by
\[{\mathcal{B}_{mj,1}} = \mathcal{B}\frac{{{d_{mj,1}}}}{{{d_{mm}}}}{\left( {\frac{{{r_{mj,1}}}}{{{r_{\max }}(z)}}} \right)^{\left| l \right|}}{e^{ - \frac{{r_{mj,1}^2 - r_{\max }^2(z)}}{{w_l^2(z)}}}}{e^{ - i\frac{{\pi (r_{mj,1}^2 - r_{\max }^2(z))}}{{\lambda {R_l}(z)}}}}{e^{ik({d_{mj,1}} - {d_{mm}})}},\tag{13}\]
where ${r_{mj,1}}$ is the radial distance between the transmission antenna ${\text{T}}{{\text{x}}_m}$ and the receive antenna ${\text{R}}{{\text{x}}_{j,1}}$. Substitute (13) into (6a), the channel response between the transmission antenna ${\text{T}}{{\text{x}}_m}$ and the receive antenna ${\text{R}}{{\text{x}}_{j,1}}$ is derived by
\[h_{mj,1}^l = \mathcal{B}\frac{\lambda }{{4\pi {d_{mm}}}}{\left( {\frac{{{r_{mj,1}}}}{{{r_{\max }}(z)}}} \right)^{\left| l \right|}}{e^{ - \frac{{r_{mj,1}^2 - r_{\max }^2(z)}}{{w_l^2(z)}}}}{e^{ - i\frac{{\pi (r_{mj,1}^2 - r_{\max }^2(z))}}{{\lambda {R_l}(z)}}}}{e^{ - ik{d_{mm}}}}{e^{ - il{\phi _{mj,1}}}}.\tag{14a}\]
Based on the derivation process in (14a), the channel response between the transmission antenna ${\text{T}}{{\text{x}}_m}$ and the receive antenna ${\text{R}}{{\text{x}}_{j,2}}$ is derived by
\[h_{mj,2}^l = \mathcal{B}\frac{\lambda }{{4\pi {d_{mm}}}}{\left( {\frac{{{r_{mj,2}}}}{{{r_{\max }}(z)}}} \right)^{\left| l \right|}}{e^{ - \frac{{r_{mj,2}^2 - r_{\max }^2(z)}}{{w_l^2(z)}}}}{e^{ - i\frac{{\pi (r_{mj,2}^2 - r_{\max }^2(z))}}{{\lambda {R_l}(z)}}}}{e^{ - ik{d_{mm}}}}{e^{ - il{\phi _{mj,2}}}}.\tag{14b}\]
In the end, the OAM wireless channel model is derived for OAM-SM millimeter wave communication systems.

\subsection{OAM Spatial Demodulation}

Assume that the perfect channel state information is obtained at receivers of OAM-SM millimeter wave communication systems. In generally, the maximum likelihood method can be adopted for demodulation of OAM-SM millimeter wave communication systems, which is expressed by
\[\left[ {\hat m,\hat l,{{\hat x}_p}} \right] = \mathop {\arg \min }\limits_{1 \leqslant m \leqslant M,l \in \mathcal{L},x \in \mathcal{X}} \left\{ {{{\left\| {{\mathbf{y}} - {\mathbf{h}}_m^l{x_p}} \right\|}^2}} \right\},\tag{15}\]
where $\hat m$ is the estimated transmission antenna index, $\hat l$ is the estimated OAM state, ${\hat x_p}$ is the estimated radiated symbol. When the OAM signal is transmitted from the $m - th$ transmission antenna, the $m - th$ receive antenna pair  are assumed to be located at the OAM circle region. In this case, the power of OAM signal at the $m - th$ receive antennas are obviously larger than the power of OAM signal at other receive antenna pairs. Hence, the transmission antenna index can be estimated by the receive antenna index with the maximum receive power of OAM signal. Based on the phase gradient method \cite{Beth36}, the OAM state is estimated by the receive antenna pair  with the maximum power of OAM signal. Furthermore, the maximum likelihood method is used to demodulate radiated symbols from the receive antenna pair. Consequently, the OAM-SM demodulation process is decomposed by the following three steps:
\[1) \quad \hat m = \mathop {\arg \;max}\limits_{1 \leqslant m \leqslant M} \left\{ {{{\left| {{{\left[ {\mathbf{y}} \right]}_{m,1}}} \right|}^2}} \right\},\tag{16}\]
\[2) \quad \hat l = \frac{{{\phi _{\hat m\hat m,1}} - {\phi _{\hat m\hat m,2}}}}{\beta },\tag{17}\]
\[3) \quad {\hat x_p} = \mathop {\arg \;\min }\limits_{x \in \mathcal{X}} \left\{ {{{\left| {{{\left[ {\mathbf{y}} \right]}_{\hat m,1}} - h_{\hat m\hat m,1}^{\hat l}{x_p}} \right|}^2}} \right\},\tag{18}\]
where ${\left[ {\mathbf{y}} \right]_{m,1}}$ is the element at the $m - th$ row and the first column of the matrix ${\mathbf{y}}$, ${\phi _{\hat m\hat m,1}}$ and ${\phi _{\hat m\hat m,2}}$ are OAM phases at the first and second antennas of the $m - th$ receive antenna pair.

\section{Capacity of OAM-SM Millimeter Wave Communication Systems}
\label{sec2}

\subsection{Capacity Model}

Assume that the radiated symbol ${x_p}$ (${x_p} \in \mathcal{X}$, $1 \leqslant p \leqslant P$) is transmitted by the $m - th$ antenna with the OAM state $l$. Based on the expression of (1), the signal received by the first antenna of the $m' - th$ ($1 \leqslant m' \leqslant M$) receive antenna pair is expressed by
\[{y_{m',1}} = \sqrt \rho  h_{mm',1}^l{x_p} + {w_{m',1}}.\tag{19}\]
Considering the assumption that ${w_{m',1}}$ is the AWGN with zero mean and variance $\sigma _{\mathbf{w}}^2$, the conditional probability density function (PDF) of ${y_{m',1}}$ is expressed by
\[\mathcal{P}\left( {{y_{m',1}}\left| {m,l,{x_p}} \right.} \right) = \frac{1}{{\pi \sigma _{\mathbf{w}}^2}}\exp \left( { - \frac{{{{\left| {{y_{m',1}} - \sqrt \rho  h_{mm',1}^l{x_p}} \right|}^2}}}{{\sigma _{\mathbf{w}}^2}}} \right).\tag{20}\]

According to the proposed OAM-SM scheme, the selections of the transmission antenna, the OAM state, and the radiated symbol are based on the random input bits. The input bits are assumed to follow independent uniform distributions. Therefore, each transmission antenna is selected with the same probability ${1 \mathord{\left/
 {\vphantom {1 M}} \right.
 \kern-\nulldelimiterspace} M}$. Similarly, each OAM state is selected with a probability ${1 \mathord{\left/
 {\vphantom {1 L}} \right.
 \kern-\nulldelimiterspace} L}$ and each radiated symbol is selected with a probability ${1 \mathord{\left/
 {\vphantom {1 P}} \right.
 \kern-\nulldelimiterspace} P}$, respectively \cite{Gallager68}. Thus, the PDF of the signal received by the first antenna of the $m' - th$ receive antenna pair, {\em i.e.}, ${y_{m',1}}$, is derived by
 \[\begin{gathered}
  \mathcal{P}\left( {{y_{m',1}}} \right) = \sum\limits_{m = 1}^M {\sum\limits_{\ell  = 1}^L {\sum\limits_{p = 1}^P {\mathcal{P}\left( {{y_{m',1}},m,l,{x_p}} \right)} } }  \hfill \\
  \;\;\;\;\;\;\;\;\;\;\;\;\;\; = \sum\limits_{m = 1}^M {\sum\limits_{\ell  = 1}^L {\sum\limits_{p = 1}^P {\mathcal{P}\left( {{y_{m',1}}\left| {m,l,{x_p}} \right.} \right)} } } \mathcal{P}\left( m \right)\mathcal{P}\left( l \right)\mathcal{P}\left( {{x_p}} \right) \hfill \\
  \;\;\;\;\;\;\;\;\;\;\;\;\;\; = \frac{1}{{MLP}}\sum\limits_{m = 1}^M {\sum\limits_{\ell  = 1}^L {\sum\limits_{p = 1}^P {\frac{1}{{\pi \sigma _{\mathbf{w}}^2}}\exp \left( { - \frac{{{{\left| {{y_{m',1}} - \sqrt \rho  h_{mm',1}^l{x_p}} \right|}^2}}}{{\sigma _{\mathbf{w}}^2}}} \right)} } }  \hfill \\
\end{gathered} .\tag{21}\]

The input symbols that are composed by the transmission antenna index, the OAM state and the radiated symbol are discrete symbols, thus they do not satisfy the requirement of Gaussian input for wireless communication systems in general. In this case, the Shannon capacity can not be achieved for wireless communication systems with non-Gaussian input \cite{Fano61}. On the other hand, because of the additive Gaussian channels in OAM-SM millimeter wave communication systems, the signals received by antenna pairs are continuous signals. Consequently, the discrete input continuous output memoryless channel (DCMC) capacity needs to be derived for analyzing the capacity of OAM-SM millimeter wave communication systems \cite{Yang08}. Furthermore, the capacity of OAM-SM millimeter wave communication systems is expressed as
\[{C_{{\text{OAM-SM}}}}{\text{ = }}\mathop {{\text{max}}}\limits_{\mathcal{P}\left( {{x_p}} \right),\mathcal{P}\left( l \right),\mathcal{P}\left( m \right)} \frac{1}{M}\sum\limits_{m' = 1}^M {\sum\limits_{m = 1}^M {\sum\limits_{\ell  = 1}^L {\sum\limits_{p = 1}^P {\int_{ - \infty }^{ + \infty } {\left\{ {\begin{array}{*{20}{c}}
  {\mathcal{P}\left( {{y_{m',1}},m,l,{x_p}} \right) \cdot } \\
  {{{\log }_2}\left( {\frac{{\mathcal{P}\left( {{y_{m',1}}\left| {m,l,{x_p}} \right.} \right)}}{{\mathcal{P}\left( {{y_{m',1}}} \right)}}} \right)}
\end{array}} \right\}} } } } {\text{d}}{y_{m',1}}} .\tag{22}\]

The mutual information between the input and output in (22) is calculated by the signal received by the first antenna of the $m' - th$ receive antenna pair, {\em i.e.}, ${y_{m',1}}$. Since the selection of the receive antenna only depends on the transmission antenna index, the mean of the mutual information corresponding to all receive antenna pairs is the average mutual information of OAM-SM millimeter wave communication systems considering all possible selections of transmission antennas. Moreover, the average mutual information in (22) is maximized when discrete inputs are uniformly distributed \cite{Goldsmith03}. In this paper discrete inputs, {\em i.e.}, the transmission antenna index, the OAM state and the radiated symbol are governed by uniform distributions. Therefore, the average mutual information in (22) is maximized and then the DCMC capacity of OAM-SM millimeter wave communication systems is obtained. Based on (20) and (21), the PDF $\mathcal{P}\left( {{y_{m',1}},m,l,{x_p}} \right)$ in (22) is derived by
\[\begin{gathered}
  \mathcal{P}\left( {{y_{m',1}},m,l,{x_p}} \right) = \mathcal{P}\left( {{y_{m',1}}\left| {m,l,{x_p}} \right.} \right)\mathcal{P}\left( {m,l,{x_p}} \right) \hfill \\
  \;\;\;\;\;\;\;\;\;\;\;\;\;\;\;\;\;\;\;\;\;\;\;\;\;\; = \frac{1}{{MLP}}\mathcal{P}\left( {{y_{m',1}}\left| {m,l,{x_p}} \right.} \right) \hfill \\
  \;\;\;\;\;\;\;\;\;\;\;\;\;\;\;\;\;\;\;\;\;\;\;\;\;\; = \frac{1}{{\pi \sigma _{\mathbf{w}}^2MLP}}\exp \left( { - \frac{{{{\left| {{y_{m',1}} - \sqrt \rho  h_{mm',1}^l{x_p}} \right|}^2}}}{{\sigma _{\mathbf{w}}^2}}} \right) \hfill \\
  \;\;\;\;\;\;\;\;\;\;\;\;\;\;\;\;\;\;\;\;\;\;\;\;\;\;\mathop  = \limits^{\left( a \right)} \frac{1}{{\pi \sigma _{\mathbf{w}}^2MLP}}\exp \left( { - \frac{{{{\left| {{w_{m',1}}} \right|}^2}}}{{\sigma _{\mathbf{w}}^2}}} \right) \hfill \\
\end{gathered} ,\tag{23}\]
where equation (a) is obtained by substituting (19) into (23). The term of ${\log _2}\left( {\frac{{\mathcal{P}\left( {{y_{m',1}}\left| {h_{mm',1}^l,{x_p}} \right.} \right)}}{{\mathcal{P}\left( {{y_{m',1}}} \right)}}} \right)$ within (22) is further derived by
\[\begin{gathered}
  {\log _2}\left( {\frac{{\mathcal{P}\left( {{y_{m',1}}\left| {h_{mm',1}^l,{x_p}} \right.} \right)}}{{\mathcal{P}\left( {{y_{m',1}}} \right)}}} \right) =  - {\log _2}\left( {\frac{{\frac{1}{{MLP}}\sum\limits_{{m_1} = 1}^M {\sum\limits_{{\ell _1} = 1}^L {\sum\limits_{{p_1} = 1}^P {\exp \left( { - \frac{{{{\left| {{y_{m',1}} - \sqrt \rho  h_{{m_1}m',1}^{{l_1}}{x_{{p_1}}}} \right|}^2}}}{{\sigma _{\mathbf{w}}^2}}} \right)} } } }}{{\exp \left( { - \frac{{{{\left| {{y_{j,1}} - \sqrt \rho  h_{mm',1}^l{x_p}} \right|}^2}}}{{\sigma _{\mathbf{w}}^2}}} \right)}}} \right) \hfill \\
  \;\;\;\;\;\;\;\;\;\;\;\;\;\;\;\;\;\;\;\;\;\;\;\;\;\;\;\;\;\;\;\;\;\;\;\;\;\;\;\;\;\;\;\; = {\log _2}MLP - {\log _2}\left( {\sum\limits_{{m_1} = 1}^M {\sum\limits_{{\ell _1} = 1}^L {\sum\limits_{{p_1} = 1}^P  \mathcal{F}\left( {{w_{m',1}}} \right)} } } \right) \hfill \\
\end{gathered} ,\tag{24}\]
where $\mathcal{F}\left( {{w_{m',1}}} \right) = \exp \left( {\frac{{{{\left| {{w_{m',1}}} \right|}^2} - {{\left| {\sqrt \rho  h_{mm',1}^l{x_p} - \sqrt \rho  h_{{m_1}m',1}^{{l_1}}{x_{{p_1}}} + {w_{m',1}}} \right|}^2}}}{{\sigma _{\mathbf{w}}^2}}} \right)$. Substitute (23) and (24) into (22), the DCMC capacity of the OAM-SM millimeter wave communication systems is finally derived by

\[\begin{gathered}
  {C_{{\text{OAM-SM}}}}{\text{ = }}\frac{1}{M}\sum\limits_{m' = 1}^M {\sum\limits_{m = 1}^M {\sum\limits_{\ell  = 1}^L {\sum\limits_{p = 1}^P {\int_{ - \infty }^{ + \infty } {\left\{ \begin{gathered}
  \frac{1}{{\pi \sigma _{\mathbf{w}}^2MLP}}\exp \left( { - \frac{{{{\left| {{w_{m',1}}} \right|}^2}}}{{\sigma _{\mathbf{w}}^2}}} \right) \cdot  \hfill \\
  \left[ {{{\log }_2}MLP - {{\log }_2}\left( {\sum\limits_{{m_1} = 1}^M {\sum\limits_{{\ell _1} = 1}^L {\sum\limits_{{p_1} = 1}^P {\mathcal{F}\left( {{w_{m',1}}} \right)} } } } \right)} \right] \hfill \\
\end{gathered}  \right\}} } } } {\text{d}}{w_{m',1}}}  \hfill \\
   \;\;\;\;\;\;\;\;\;\;\;\;\;\; = \frac{1}{M}\sum\limits_{m' = 1}^M {\left\{ {{{\log }_2}MLP - \frac{1}{{MLP}}\sum\limits_{m = 1}^M {\sum\limits_{\ell  = 1}^L {\sum\limits_{p = 1}^P {{\mathbb{E}_{{w_{m',1}}}}\left[ {{{\log }_2}\left( {\sum\limits_{{m_1} = 1}^M {\sum\limits_{{\ell _1} = 1}^L {\sum\limits_{{p_1} = 1}^P {\mathcal{F}\left( {{w_{m',1}}} \right)} } } } \right)} \right]} } } } \right\}}  \hfill \\
\end{gathered} ,\tag{25}\]
where ${\mathbb{E}_{{w_{m',1}}}}$ stands for the expectation operation with respect to ${w_{m',1}}$.

\subsection{Performance Analysis of Capacity}

Based on the proposed capacity model of OAM-SM millimeter wave communication systems, the performance of capacity is simulated. Moreover, a conventional MIMO communication system \cite{Goldsmith03} are simulated and compared with the proposed OAM-SM millimeter wave communication systems in the following performance analysis.
The default simulation parameters are configured as follows: the carrier frequency is 60 GHz, the bandwidth is 20 MHz, the OAM state is 8, the transmission distance between the transmitter and receiver is $d = 50$ meters, the noise power spectral density is -174 dBm/Hz \cite{Mao13,Mao06}, and the distance between transmission (receive) antenna elements is $\xi  = 20\lambda $ \cite{Hui15,Zhang16,Xiang13}. The OAM-SM communication systems are configured with 4 transmission antennas and 4 receive antenna pairs. The MIMO communication systems are configured with 4 transmission antennas and 4 receive antennas. The 4PSK modulation is adopted for OAM-SM and MIMO communication systems. The total transmit power of the MIMO communication system is configured to be equal to the transmit power of OAM-SM communication systems.

Fig. 2 illustrates the capacity of OAM-SM and MIMO millimeter wave communication systems with respect to the transmission SNR. As shown in Fig. 2, it is observed that the capacity of OAM-SM and MIMO millimeter wave communication systems always increases with the increase of the transmission SNR. When the transmission SNR is fixed, the capacity of OAM-SM millimeter wave communication system is larger than the capacity of MIMO millimeter wave communication system. Compared with the capacity of MIMO millimeter wave communication systems, the maximum capacity of OAM-SM millimeter wave communication system is improved by 36\%.

In Fig.3, the capacity with respect to the transmission distance between the transmitter and the receiver is analyzed for OAM-SM and MIMO millimeter wave communication systems. From Fig. 3, the capacity of OAM-SM and MIMO millimeter wave communication systems decreases with the increase of the transmission distance. When the transmission distance is less than 38 meters, the capacity of OAM-SM millimeter wave communication systems is less than the capacity of MIMO millimeter wave communication systems. When the transmission distance is larger than or equal to 38 meters, the capacity of OAM-SM millimeter wave communication systems is larger than the capacity of MIMO millimeter wave communication systems. Hence, OAM-SM millimeter wave communication systems are more suitable for wireless transmission in a long distance than conventional MIMO millimeter wave communication systems.

The capacity of OAM-SM millimeter wave communication systems with respect to the transmission SNR considering different numbers of transmission antennas is illustrated in Fig. 4(a). When the transmission SNR is less than -15 dB, the capacity of OAM-SM millimeter wave communication systems decreases with the increase of the number of transmission antennas. When the transmission SNR is larger than or equal to -15 dB, the capacity of OAM-SM millimeter wave communication systems increases with the increase of the number of transmission antennas. The capacity of OAM-SM millimeter wave communication systems with respect to the transmission SNR considering different numbers of OAM states is illustrated in Fig. 4(b). When the transmission SNR is fixed, the capacity of OAM-SM millimeter wave communication systems decreases with the increase of the number of OAM states.

The capacity of OAM-SM millimeter wave communication systems with respect to the transmission SNR considering different $P$-point constellation modulation methods is illustrated in Fig. 5. When $P$ is configured as 2, 4, and 8, $P$-point constellation modulation corresponds to 2PSK, 4PSK and 8PSK, respectively. When the transmission SNR is less than -15 dB, the capacity of OAM-SM millimeter wave communication systems decreases with the increase of the point number in $P$-point constellation modulation methods. When the transmission SNR is larger than or equal to -15 dB, the capacity of OAM-SM millimeter wave communication systems increases with the increase of the point number in $P$-point constellation modulation methods.

The capacity of OAM-SM millimeter wave communication systems with respect to the transmission distance considering different numbers of transmitting antennas is shown in Fig. 6(a). When the transmission distance is fixed, the capacity of OAM-SM millimeter wave communication systems increases with the increase of the number of transmission antennas. The capacity of OAM-SM millimeter wave communication systems with respect to the transmission distance considering different numbers of OAM states is presented at Fig. 6(b). When the transmission distance is less than 17 meters, the capacity of OAM-SM millimeter wave communication systems increases with the increase of the number of OAM states. When the transmission distance is larger than or equal to 17 meters, the capacity of OAM-SM millimeter wave communication systems decreases with the increase of the number of OAM states.

The capacity of OAM-SM millimeter wave communication systems with respect to the transmission distance considering different $P$-point constellation modulation methods is depicted in Fig. 7. When the transmission distance is fixed, the capacity of OAM-SM millimeter wave communication systems increases with the increase of the point number in $P$-point constellation modulation methods.

\section{Average Bit Error Probability of OAM-SM Millimeter Wave Communication Systems}
\label{sec2}

\subsection{Average Bit Error Probability Models}

Based on the proposed OAM-SM millimeter wave communication systems, the receiver retrieves symbols conveyed from receive signals, which is demodulated from the transmission antenna index, the OAM state and the radiated symbol. Assume that correct demodulation events conveyed from the transmission antenna index, the OAM state and the radiated symbol are denoted as ${\Delta _{{\text{ant}}}}$, ${\Delta _{{\text{OAM}}}}$, ${\Delta _{{\text{mod}}}}$, respectively. If there exist any errors in correct demodulation events, the total demodulation process of OAM-SM millimeter wave communication systems is determined as a failure. Therefore, the probability of the total correct demodulation event $\Delta $ is expressed by
\[\begin{gathered}
  \mathcal{P}\left( \Delta  \right) = \mathcal{P}\left( {{\Delta _{{\text{ant}}}}{\Delta _{{\text{OAM}}}}{\Delta _{{\text{mod}}}}} \right) \hfill \\
   \;\;\;\;\;\;\;\;\;\; = \mathcal{P}\left( {{\Delta _{{\text{mod}}}}\left| {{\Delta _{{\text{ant}}}}{\Delta _{{\text{OAM}}}}} \right.} \right)\mathcal{P}\left( {{\Delta _{{\text{OAM}}}}\left| {{\Delta _{{\text{ant}}}}} \right.} \right)\mathcal{P}\left( {{\Delta _{{\text{ant}}}}} \right) \hfill \\
\end{gathered} ,\tag{26}\]
where $\mathcal{P}\left(  \cdot  \right)$ is the probability of an event, $\mathcal{P}\left( {{\Delta _{{\text{mod}}}}\left| {{\Delta _{{\text{ant}}}}{\Delta _{{\text{OAM}}}}} \right.} \right)$ is the conditional correct estimation probability of bits conveyed by the radiated symbol while bits conveyed by the transmission antenna index and the OAM state have already been correctly demodulated. $\mathcal{P}\left( {{\Delta _{{\text{OAM}}}}\left| {{\Delta _{{\text{ant}}}}} \right.} \right)$ is the conditional correct estimation probability of bits conveyed by the OAM state while bits conveyed by the transmission antenna index have already been correctly demodulated. $\mathcal{P}\left( {{\Delta _{{\text{ant}}}}} \right)$ is the correct estimation probability of bits conveyed by the transmission antenna index. Based on (26), the incorrect demodulation probability, {\em i.e.} the ABEP is expressed by
\[\begin{gathered}
  \mathcal{P}\left( \mathcal{E} \right) = 1 - \mathcal{P}\left( \Delta  \right) \hfill \\
  \;\;\;\;\;\;\;\;\; = 1 - \mathcal{P}\left( {{\Delta _{{\text{mod}}}}\left| {{\Delta _{{\text{ant}}}}{\Delta _{{\text{OAM}}}}} \right.} \right)\mathcal{P}\left( {{\Delta _{{\text{OAM}}}}\left| {{\Delta _{{\text{ant}}}}} \right.} \right)\mathcal{P}\left( {{\Delta _{{\text{ant}}}}} \right) \hfill \\
\end{gathered} .\tag{27}\]

In the proposed OAM-SM millimeter wave communication systems, the judgement of the transmission antenna index depends on the power strength of receive antenna pairs. Based on OAM-SM scheme, the receive antenna pair corresponding to the transmission antenna is configured within the intensity-focused circle region where the transmitted OAM beam has the maximum intensity. Since two antennas of a receive antenna pair corresponding to the transmission antenna locate within the same intensity-focused circle region, the received power of two antennas of a receive antenna pair is equal. In this case, only one of two antennas of receive antenna pair is used when the signal power of receive antenna pairs needs to be compared. Without loss of generality, the $m - th$ transmission antenna ${\text{T}}{{\text{x}}_m}$ is assumed for transmitting OAM signals in OAM-SM millimeter wave communication systems. The corresponding receive antenna ${\text{R}}{{\text{x}}_{m,1}}$ is located within the intensity-focused circle region while the other receive antenna ${\text{R}}{{\text{x}}_{j,1}}$ ($j \ne m$) is located outside the intensity-focused circle region. The signal received at the antenna ${\text{R}}{{\text{x}}_{m,1}}$ is expressed as
\[{y_{m,1}} = \sqrt \rho  h_{mm}^l{x_p} + {w_{m,1}},\tag{28}\]
where the AWGN is governed by ${w_{m,1}} \sim \mathcal{C}\mathcal{N}\left( {0,\sigma _w^2} \right)$. Let $\sigma _0^2 = {{\sigma _w^2} \mathord{\left/
 {\vphantom {{\sigma _w^2} 2}} \right.
 \kern-\nulldelimiterspace} 2}$, the following results are derived by
 \[\begin{gathered}
  {\left| {{y_{m,1}}} \right|^2} = \mathbb{R}{({y_{m,1}})^2} + \mathbb{I}{({y_{m,1}})^2} \hfill \\
  \quad \quad  \sim \mathcal{N}\left( {\mathbb{R}\left( {\sqrt \rho  h_{mm}^l{x_p}} \right),\sigma _0^2} \right) + \mathcal{N}\left( {\mathbb{I}\left( {\sqrt \rho  h_{mm}^l{x_p}} \right),\sigma _0^2} \right) \hfill \\
\end{gathered},\tag{29a} \]
\[\begin{gathered}
  {\left| {{y_{j,1}}} \right|^2} = \mathbb{R}{({y_{j,1}})^2} + \mathbb{I}{({y_{j,1}})^2} \hfill \\
  \quad \quad  \sim \mathcal{N}\left( {\mathbb{R}\left( {\sqrt \rho  h_{mj}^l{x_p}} \right),\sigma _0^2} \right) + \mathcal{N}\left( {\mathbb{I}\left( {\sqrt \rho  h_{mj}^l{x_p}} \right),\sigma _0^2} \right) \hfill \\
\end{gathered},\tag{29b} \]
where $\mathbb{R}\left(  \cdot  \right)$ and $\mathbb{I}\left(  \cdot  \right)$ denote operations taking the real and imaginary part from a complex variable, respectively. Based on the definition of chi-square distribution, (29) is further formulated by \cite{Mood74}
\[{\left| {{y_{m,1}}} \right|^2} \sim \chi _2^2\left( {g;{\lambda _{m,1}}} \right),\tag{30a}\]
\[{\left| {{y_{j,1}}} \right|^2} \sim \chi _2^2\left( {g;{\lambda _{j,1}}} \right),\tag{30b}\]
where $\chi _2^2\left( {g;{\lambda _{m,1}}} \right)$ is a chi-square distribution with the random variable $g$, two degrees of freedom and non-centrality ${\lambda _{m,1}}$. In (30) the non-centralities are expressed as ${\lambda _{m,1}} = {{{{\left| {\rho h_{mm}^l{x_p}} \right|}^2}} \mathord{\left/
 {\vphantom {{{{\left| {\rho h_{mm}^l{x_p}} \right|}^2}} {\sigma _0^2}}} \right.
 \kern-\nulldelimiterspace} {\sigma _0^2}}$ and ${\lambda _{j,1}} = {{{{\left| {\rho h_{mj}^l{x_p}} \right|}^2}} \mathord{\left/
 {\vphantom {{{{\left| {\rho h_{mj}^l{x_p}} \right|}^2}} {\sigma _0^2}}} \right.
 \kern-\nulldelimiterspace} {\sigma _0^2}}$. Based on the power strength of received signals, the probability that the transmission antenna index is correctly estimated is derived by
\[\begin{gathered}
  \mathcal{P}\left( {{\Theta _{{\text{ant}}}}\left| {{\lambda _{m,1}}} \right.} \right) = \int_0^\infty  {\left[ {\mathcal{P}\left( {{{\left| {{y_{1,1}}} \right|}^2} < {g_{m,1}}, \cdots ,{{\left| {{y_{m - 1,1}}} \right|}^2} < {g_{m,1}},{{\left| {{y_{m + 1,1}}} \right|}^2} < {g_{m,1}}, \cdots ,} \right.} \right.}  \hfill \\
  \;\;\;\;\;\;\;\;\;\;\;\;\;\;\;\;\;\;\;\;\;\;\;\; \left. {\left. {{{\left| {{y_{M,1}}} \right|}^2} < {g_{m,1}}} \right) \cdot \mathcal{P}\left( {{{\left| {{y_{m,1}}} \right|}^2} = {g_{m,1}}\left| {{\lambda _{m,1}}} \right.} \right)} \right]d{g_{m,1}} \hfill \\
  \;\;\;\;\;\;\;\;\;\;\;\;\;\;\;\;\;\;\;\;\; \mathop  = \limits^{(a)} \int_0^\infty  {\prod\limits_{j = 1,j \ne m}^M {\mathcal{P}\left( {{{\left| {{y_{j,1}}} \right|}^2} < {g_{m,1}}} \right)} } \mathcal{P}\left( {{{\left| {{y_{m,1}}} \right|}^2} = {g_{m,1}}\left| {{\lambda _{m,1}}} \right.} \right)d{g_{m,1}} \hfill \\
  \;\;\;\;\;\;\;\;\;\;\;\;\;\;\;\;\;\;\;\;\; = \int\limits_0^\infty  {\left[ {\prod\limits_{j = 1,j \ne m}^M {{F_{\chi _2^2}}\left( {g;{\lambda _{j,1}}} \right)} } \right]}  \cdot {f_{\chi _2^2}}\left( {g;{\lambda _{m,1}}} \right)dg \hfill \\
\end{gathered},\tag{31} \]
where ${\Theta _{{\text{ant}}}}$ is the event that the transmission antenna index is correctly estimated. $\mathcal{P}\left( {{{\left| {{y_{m,1}}} \right|}^2} = {g_{m,1}}\left| {{\lambda _{m,1}}} \right.} \right)$ is the probability that the amplitude of the signal received by the antenna ${\text{R}}{{\text{x}}_{m,1}}$ equals to ${g_{m,1}}$ when the non-centrality ${\lambda _{m,1}}$ is given. $\mathcal{P}\left( {{{\left| {{y_{1,1}}} \right|}^2} < {g_{m,1}}, \cdots ,{{\left| {{y_{M,1}}} \right|}^2} < {g_{m,1}}} \right)$ is the probability that the power of the signals received by the other antennas ${\text{R}}{{\text{x}}_{j,1}}$ ($j \ne m$) are less than ${g_{m,1}}$. ${F_{\chi _2^2}}\left( {g;{\lambda _{j,1}}} \right)$ is the cumulative distribution function (CDF) of the chi-square distribution with the random variable $g$, two degrees of freedom and non-centrality ${\lambda _{j,1}}$. ${f_{\chi _2^2}}\left( {g;{\lambda _{m,1}}} \right)$ is the PDF of the chi-square distribution with the random variable $g$, two degrees of freedom and non-centrality ${\lambda _{m,1}}$. Equation (a) within (31) is established with the constraint that signals received by different receive antennas are independence each other \cite{Wang11}. Furthermore, the average probability that the transmission antenna index is correctly estimated is derived by
\[\mathcal{P}\left( {{\Theta _{{\text{ant}}}}} \right) = \frac{1}{M}\sum\limits_{m = 1}^M {\mathcal{P}\left( {{\Delta _{{\text{ant}}}}\left| {{\lambda _{m,1}}} \right.} \right)}  = \frac{1}{M}\sum\limits_{m = 1}^M {\int\limits_0^\infty  {\left[ {\prod\limits_{j = 1,j \ne m}^M {{F_{\chi _2^2}}\left( {g;{\lambda _{j,1}}} \right)} } \right]}  \cdot {f_{\chi _2^2}}\left( {g;{\lambda _{m,1}}} \right)dg} .\tag{32}\]

When the transmission antenna index is not correctly estimated, the average symbol error probability of the transmission antenna index is expressed by
\[e_{{\text{ant}}}^{\text{s}} = 1 - \mathcal{P}\left( {{\Theta _{{\text{ant}}}}} \right).\tag{33}\]
Furthermore, the ABEP conveyed by the transmission antenna index $e_{{\text{ant}}}^{\text{b}}$ is expressed by \cite{Zhang15}
\[e_{{\text{ant}}}^{\text{b}} = \frac{{e_{{\text{ant}}}^{\text{s}} \cdot {\gamma _{{{\log }_2}M}}}}{{{{\log }_2}M}},\tag{34}\]
where ${\gamma _{{{\log }_2}M}}$ is extended by ${\gamma _{{{\log }_2}M}} = {\gamma _{{{\log }_2}M - 1}} + \frac{{{2^{{{\log }_2}M - 1}} - {\gamma _{{{\log }_2}M - 1}}}}{{{2^{{{\log }_2}M}} - 1}}$ with an initial condition ${\gamma _0} = 0$. Consequently, the correct estimation probability of bits conveyed by the transmit antenna index is derived by
\[\mathcal{P}\left( {{\Delta _{{\text{ant}}}}} \right) = 1 - e_{{\text{ant}}}^{\text{b}}.\tag{35}\]

When the transmission antenna ${\text{T}}{{\text{x}}_m}$ has been estimated, OAM phases at the corresponding receive antenna pair are denoted as ${\phi _{mm,1}}$ and ${\phi _{mm,2}}$, respectively. As a consequence, the OAM state of the receive signal is derived by \cite{Mohammadi10}
\[l = \frac{{{\phi _{mm,1}} - {\phi _{mm,2}}}}{\beta }.\tag{36}\]

As illustrated in Fig. 1, the azimuthal angle $\beta $ between ${\text{R}}{{\text{x}}_{m,1}}$ and ${\text{R}}{{\text{x}}_{m,2}}$ should satisfy $\beta  < \frac{\pi }{{\left| {{L_{\max }}} \right|}}$. In this case, the OAM state $l$ is always correctly estimated \cite{Allen14,Mohammadi10}. In other words, the corresponding receive antennas always obtain the correct OAM state based on (36) if the transmit antenna index is correctly estimated. Therefore, we have
\[\mathcal{P}\left( {{\Delta _{{\text{OAM}}}}\left| {{\Delta _{{\text{ant}}}}} \right.} \right) = 1.\tag{37}\]

Based on the OAM-SM scheme, the maximum likelihood method is adopted to demodulate the radiated symbol after the transmission antenna index and the OAM state have been estimated. Hence, the bit error probability of the radiated symbol is derived by \cite{Irshid91,Simon05}
\[e_{{\text{mod}}}^{\text{b}} = \frac{1}{{P{{\log }_2}P}}\sum\limits_{p = 1}^P {\sum\limits_{{p_1} = 1}^P {\mathbb{D}({x_p},{x_{{p_1}}})} \mathbb{Q}\left( {\sqrt {\frac{\rho }{{2\sigma _w^2}}{{\left| {h_{mm}^l} \right|}^2}{{\left| {{x_p},{x_{{p_1}}}} \right|}^2}} } \right)} ,\tag{38}\]
where ${x_p}$ is the input radiated symbol and ${x_{{p_1}}}$ is the demodulated output radiated symbol, $\mathbb{D}({x_p},{x_{{p_1}}})$ is the Hamming distance between ${x_p}$ and ${x_{{p_1}}}$, $\mathbb{Q}\left(  \cdot  \right)$ is the Q-function \cite{Simon05}, $\left| {{x_p},{x_{{p_1}}}} \right| = 2\mathbb{R}\left[ {{x_p}{{\left( {{x_p} - {x_{{p_1}}}} \right)}^ * }} \right]$ is the codeword difference between ${x_p}$ and ${x_{{p_1}}}$. Furthermore, the conditional correct estimation probability of bits conveyed by the radiated symbol is derived by
\[\mathcal{P}\left( {{\Delta _{{\text{mod}}}}\left| {{\Delta _{{\text{ant}}}}{\Delta _{{\text{OAM}}}}} \right.} \right) = 1 - e_{{\text{mod}}}^{\text{b}}.\tag{39}\]

Substitute (32)-(35), (37)-(39) into (27), the ABEP of the OAM-SM millimeter wave communication systems is derived by

\[\begin{gathered}
  \mathcal{P}\left( \mathcal{E} \right) = 1 - \left[ {1 - \frac{{\left( {1 - \frac{1}{M}\sum\limits_{m = 1}^M {\int\limits_0^\infty  {\left[ {\prod\limits_{j = 1,j \ne m}^M {{F_{\chi _2^2}}\left( {g;{\lambda _{j,1}}} \right)} } \right]}  \cdot {f_{\chi _2^2}}\left( {g;{\lambda _{m,1}}} \right)dg} } \right) \cdot {\gamma _{{{\log }_2}M}}}}{{{{\log }_2}M}}} \right] \hfill \\
  \;\;\;\;\;\;\;\;\;\;\;\;\; \cdot \left[ {1 - \frac{1}{{P{{\log }_2}P}}\sum\limits_{p = 1}^P {\sum\limits_{{p_1} = 1}^P {\mathbb{D}({x_p},{x_{{p_1}}})} \mathbb{Q}\left( {\sqrt {\frac{\rho }{{2\sigma _w^2}}{{\left| {h_{mm}^l} \right|}^2}{{\left| {{x_p},{x_{{p_1}}}} \right|}^2}} } \right)} } \right] \hfill \\
\end{gathered} .\tag{40}\]

\subsection{Performance Analysis of Average Bit Error Probability}

To analyze the performance of ABEP in OAM-SM millimeter wave communication systems, some default parameters are configured for simulations as follows: the number of transmission antennas is 4, the 4PSK modulation is adopted, the number of OAM state is 8, the distance of transmitter and receiver is 50 meters.

Fig. 8 illustrates the ABEP of OAM-SM and MIMO millimeter wave communication systems with respect to the transmission SNR. When the transmission SNR is less than 7 dB, the ABEP of MIMO millimeter wave communication systems is larger than or equal to the ABEP of OAM-SM millimeter wave communication systems. When the transmission SNR is in the range of 7 dB and 19 dB, the ABEP of OAM-SM millimeter wave communication systems is larger than the ABEP of MIMO millimeter wave communication systems. When the transmission SNR is larger than 19 dB, the ABEP of MIMO millimeter wave communication systems is larger than or equal to the ABEP of OAM-SM millimeter wave communication systems.

Fig. 9(a) shows the ABEP of OAM-SM millimeter wave communication systems with respect to the transmission SNR considering different numbers of transmission antennas. When the number of transmission antennas is fixed, the ABEP of OAM-SM millimeter wave communication systems decreases with the increase of the transmission SNR. When the transmission SNR is fixed, the ABEP of OAM-SM millimeter wave communication systems increases with the increase of the number of transmission antennas. When the transmission SNR is less than 10 dB, the impact of the number of transmission antennas on the ABEP is obviously reduced. The ABEP of OAM-SM millimeter wave communication systems with respect to the transmission SNR considering different P-point constellation modulation schemes is illustrated in Fig 9(b). When the transmission SNR is fixed, the ABEP of OAM-SM millimeter wave communication systems increases with the increase of the point number in $P$-point constellation modulation schemes. When the transmission SNR is larger than 10 dB, the impact of different $P$-point constellation modulation schemes on the ABEP is obviously reduced.

Fig. 10(a) presents the ABEP of OAM-SM millimeter wave communication systems with respect to the transmission SNR considering different distances between transmitters and receivers. When the transmission SNR is fixed, the ABEP of OAM-SM millimeter wave communication systems increases with the increase of the distance between transmitters and receivers. The ABEP of OAM-SM millimeter wave communication systems with respect to the transmission SNR considering different numbers of OAM states is depicted in Fig. 10(b). When the transmission SNR is fixed, the ABEP of OAM-SM millimeter wave communication systems increases with the increase of the number of OAM states.

\section{Energy Efficiency of OAM-SM Communication Systems}
\label{sec2}

\subsection{Energy Efficiency Models}

Considering the circuit energy of communication systems, the energy efficiency of OAM-SM millimeter wave communication systems is analyzed and compared with the energy efficiency of conventional MIMO communication systems in the following. Assuming the transmit power is equally distributed among all transmission antennas, the power consumed by the multi-antenna transmitter is modelled as \cite{Simon05}
\[{\rho ^{{\text{total}}}} = {N_{\text{C}}}{\rho _{\text{C}}} + \alpha {N_{\text{C}}}\rho ,\tag{41}\]
where ${N_{\text{C}}}$ is the number of the active transmission antennas, ${\rho _{\text{C}}}$ is the power consumed by circuits associated with an active transmission antenna, $\rho$ is the transmit power of a transmission antenna, $\alpha$ is the slope of the load dependent power consumption. Comparing to transmitters, the power consumed by the receiver is very small and negligible \cite{Simon05,Stavridis12}. Therefore, only the power consumed by transmitters is considered for OAM-SM and MIMO communication systems. In general, the energy efficiency of communication systems is expressed by
\[\eta  = \frac{{W \cdot C\left( {{N_{\text{C}}}\rho } \right)}}{{{\rho ^{{\text{total}}}}}},\tag{42}\]
where ${N_{\text{C}}}\rho$ is the total transmit power, $W$ denotes the bandwidth, $C\left( {{N_{\text{C}}}\rho } \right)$ denotes the capacity as a function of the total transmit power. In OAM-SM communication systems, at any time only a single transmission antenna is active. Consequently, ${N_{\text{C}}}$ is configured as one for OAM-SM communication systems. In this case, the energy efficiency of OAM-SM communication system is expressed by
\[{\eta _{{\text{OAM - SM}}}} = \frac{{W \cdot {C_{{\text{OAM - SM}}}}\left( \rho  \right)}}{{{\rho _{\text{C}}} + \alpha {\rho _{\text{t}}}}},\tag{43}\]
where ${C_{{\text{OAM - SM}}}}\left( \rho  \right)$ is derived by (25).

For MIMO communication systems with $M$ transmission antennas and $M$ receive antennas, the energy efficiency is derived by
\[{\eta _{{\text{MIMO}}}} = \frac{{W \cdot {C_{{\text{MIMO}}}}\left( {M{\rho ^{{\text{MIMO}}}}} \right)}}{{M{\rho _{\text{C}}} + \alpha M\rho _{\text{t}}^{{\text{MIMO}}}}},\tag{44}\]
where the capacity ${C_{{\text{MIMO}}}}\left( {M{\rho ^{{\text{MIMO}}}}} \right)$ is given by \cite[Eq.8]{Serafimovski13}, $\rho ^{{\text{MIMO}}}$ is the transmission power at every antenna.

\subsection{Performance Analysis of Energy Efficiency}

To evaluate the energy efficiency of OAM-SM and MIMO millimeter wave communication systems, some default parameters are configured as follows: the power consumed by circuits associated with each active transmission antenna, {\em i.e.}, ${\rho _{\text{C}}}$ is 6.8 W, the slope of the load dependent power consumption, {\em i.e.}, $\alpha $ is configured as 4.0 \cite{Auer11}, the distance of transmitter and receiver is 50 meters. To compare the OAM-SM millimeter wave communication systems with MIMO millimeter wave communication systems, the transmit power of OAM-SM millimeter wave communication systems is configured to be equal to the transmit power of MIMO millimeter wave communication systems, {\em i.e.}, $\rho  = M{\rho ^{{\text{MIMO}}}}$.

The energy efficiency of OAM-SM and MIMO millimeter wave communication systems with respect to the transmission SNR is illustrated in Fig. 11. When the transmission SNR is less than 10 dB, the energy efficiency of OAM-SM millimeter wave communication systems increase with the increase of the transmission SNR. When the transmission SNR is larger than or equal to 10 dB, the energy efficiency of OAM-SM millimeter wave communication systems decrease with the increase of the transmission SNR. There exist a maximum energy efficiency of OAM-SM millimeter wave communication systems considering different transmission SNR values. When the transmission SNR is fixed, the energy efficiency of OAM-SM millimeter wave communication systems is always larger than the energy efficiency of MIMO millimeter wave communication systems. To be specific, compared with the energy efficiency of MIMO millimeter wave communication systems, the maximum energy efficiency of OAM-SM millimeter wave communication system is improved by 472.3\%.

\section{Conclusions}
\label{sec2}

Exploiting spatial distribution characteristics of OAM beams, a new OAM-SM millimeter wave communication system is first proposed for future mobile networks. Important performance metrics like capacity, ABEP and energy efficiency of OAM-SM millimeter wave communication systems are analytically obtained in closed forms. It is shown analytically that significant performance gains can be achieved by the OAM-SM millimeter wave communication systems over the conventional MIMO millimeter wave communication systems, e.g., the maximum capacity and energy efficiency are improved by 36\% and 472.3\%, respectively. Furthermore, numerical results indicate that the OAM-SM millimeter wave communication system is more suitable for long-range communications, where a higher capacity is achieved than the conventional MIMO millimeter wave communication system when the distance is larger than 38 meters. Owing to its promising performance, the proposed OAM-SM millimeter wave communication system can be regarded as a candidate solution to future mobile networks. It is a challenge to perfectly align the transmission and receive antenna arrays in the implementation of OAM-SM millimeter wave communication systems. For the future work, we will explore potential solutions to OAM-SM millimeter wave communication systems with misaligned antenna arrays in future mobile networks.

\newpage

\begin{figure}
\centering
\includegraphics[width=15cm]{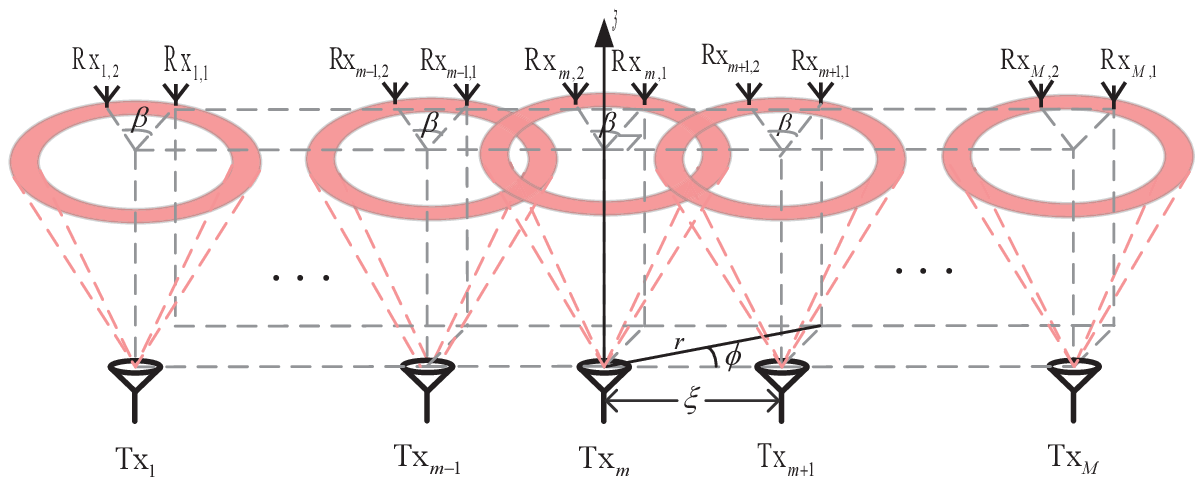}
\begin{quote}
\small Fig.1. System model of OAM-SM millimeter wave communications
\end{quote}
\end{figure}

\begin{figure}
\centering
\includegraphics[width=10cm,draft=false]{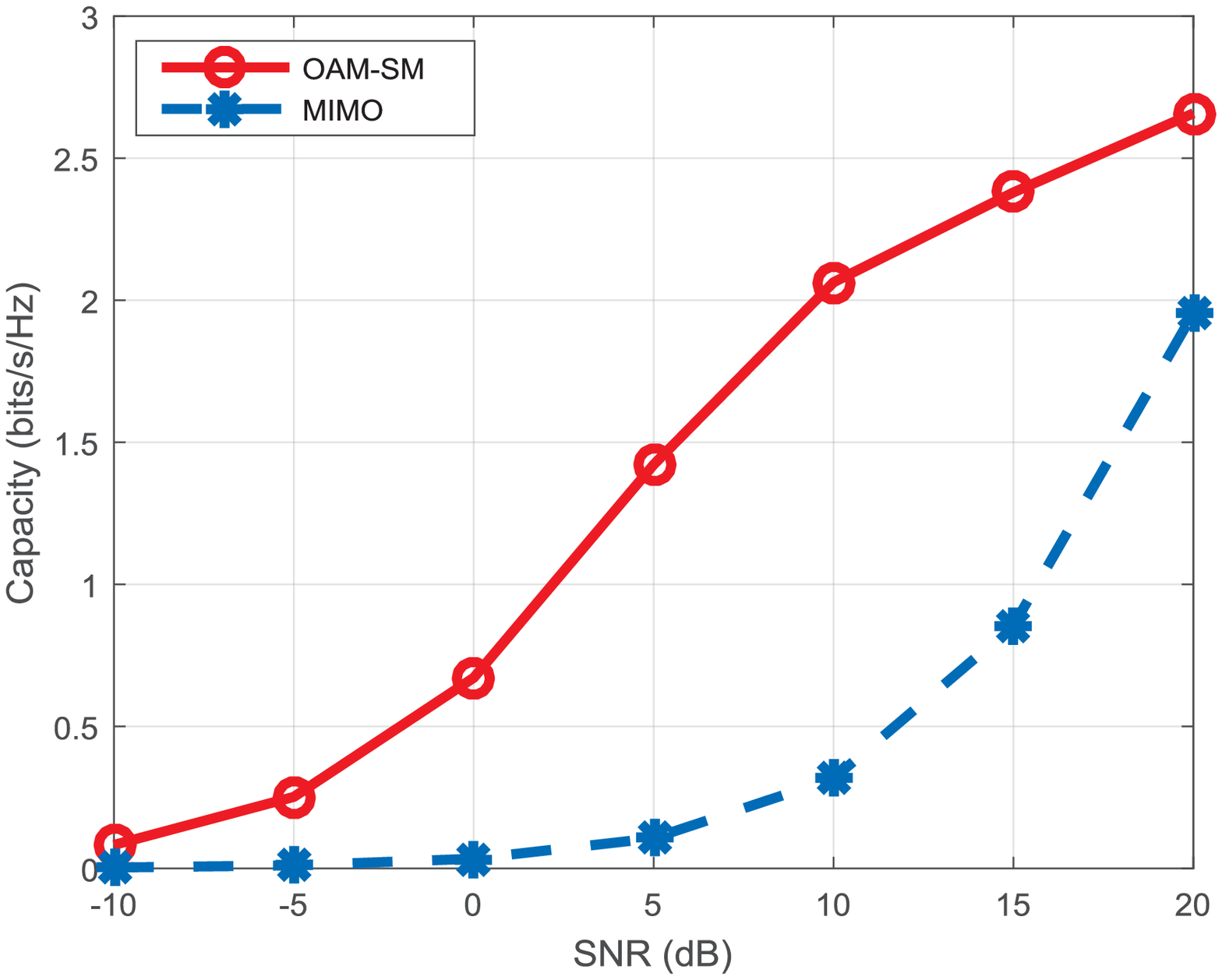}
\begin{quote}
\small Fig.2. Capacity of OAM-SM and MIMO millimeter wave communication systems with respect to the transmission SNR.
\end{quote}
\end{figure}

\begin{figure}
\centering
\includegraphics[width=10cm,draft=false]{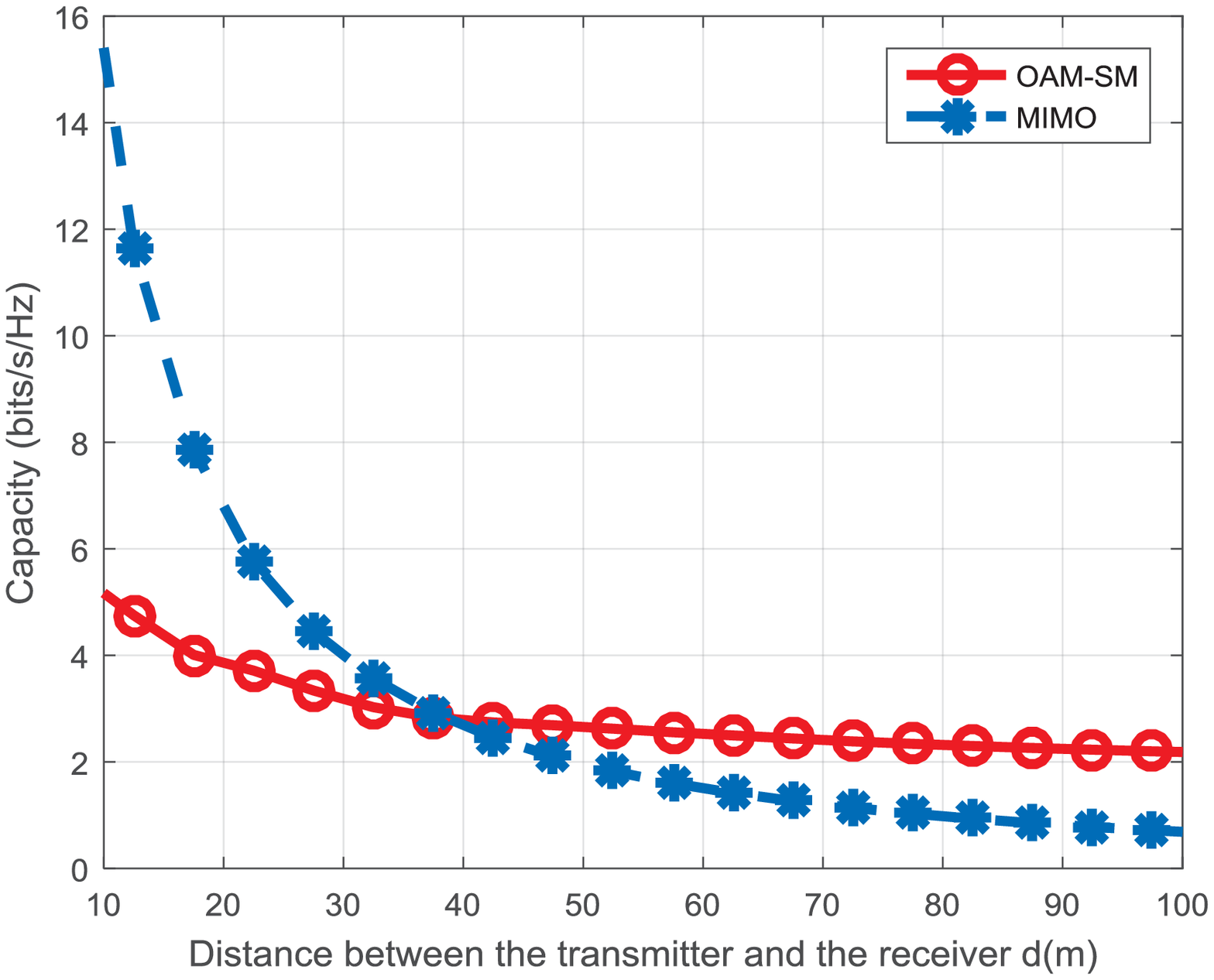}
\begin{quote}
\small Fig.3. Capacity of OAM-SM and MIMO millimeter wave communication systems with respect to the distance between the transmitter and the receiver.
\end{quote}
\end{figure}

\begin{figure}
\centering
\includegraphics[width=15cm,draft=false]{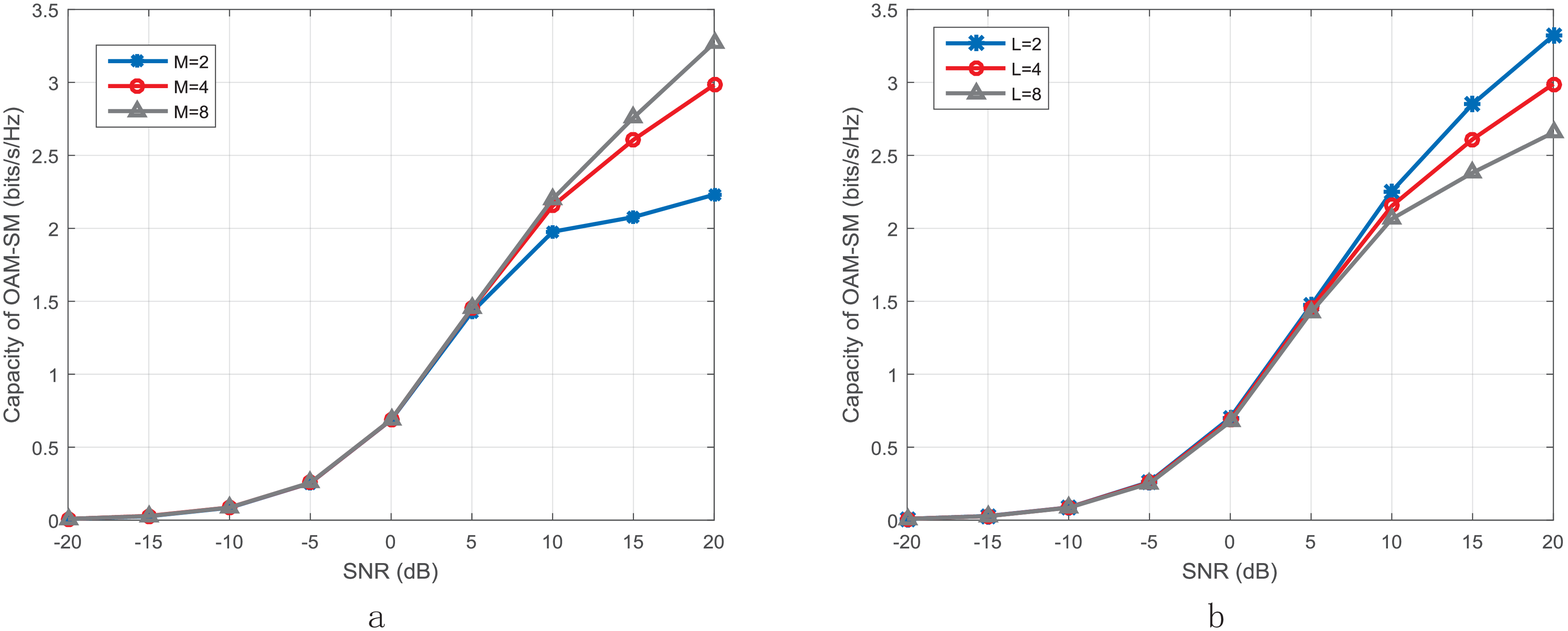}
\begin{quote}
\small Fig.4. Capacity of OAM-SM millimeter wave communication systems with respect to the transmission SNR considering different number of transmission antennas and OAM states.
\end{quote}
\end{figure}

\begin{figure}
\centering
\includegraphics[width=10cm,draft=false]{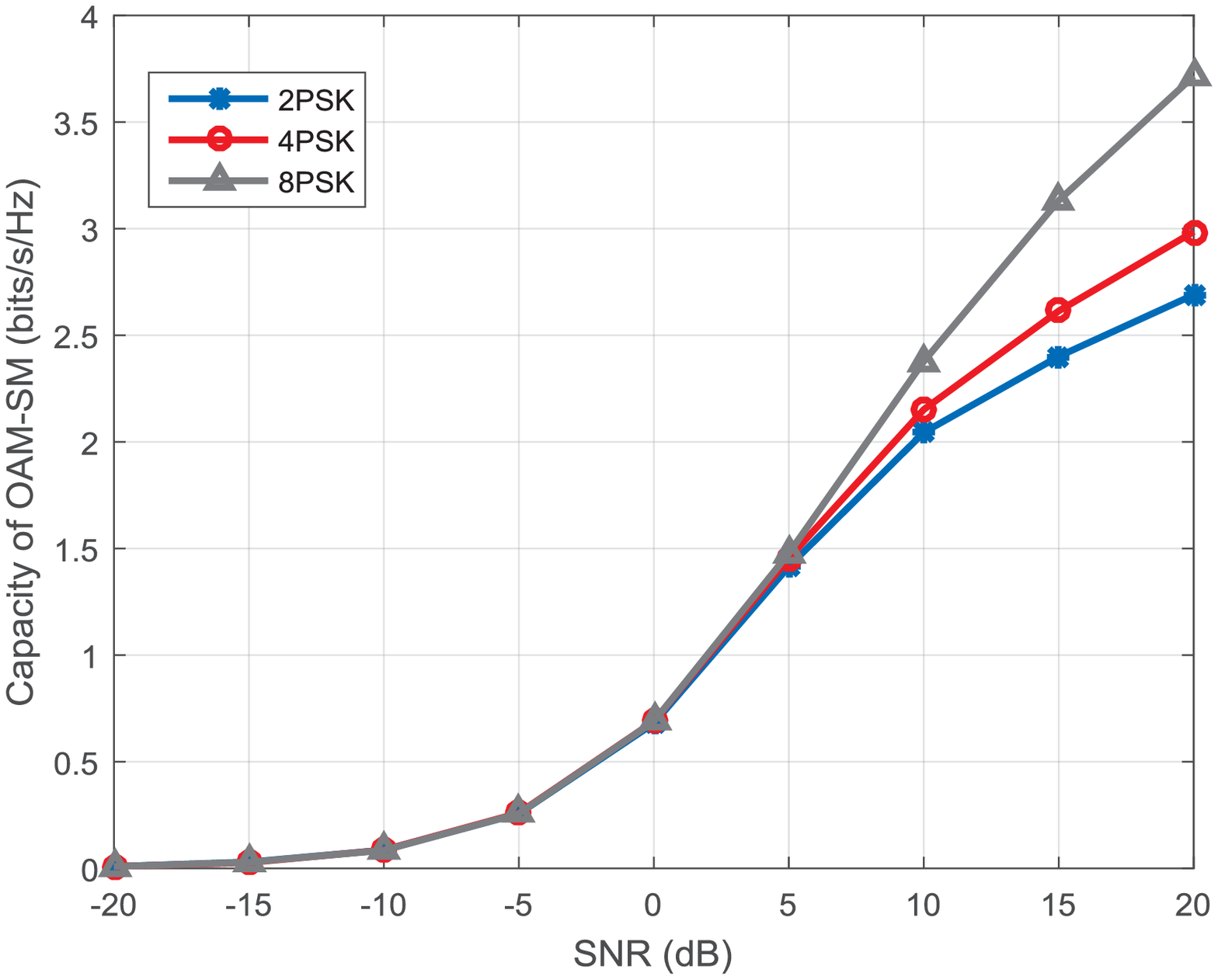}
\begin{quote}
\small Fig.5. Capacity of OAM-SM millimeter wave communication systems with respect to the transmission SNR considering different $P$-point constellation modulation methods.
\end{quote}
\end{figure}

\begin{figure}
\centering
\includegraphics[width=15cm,draft=false]{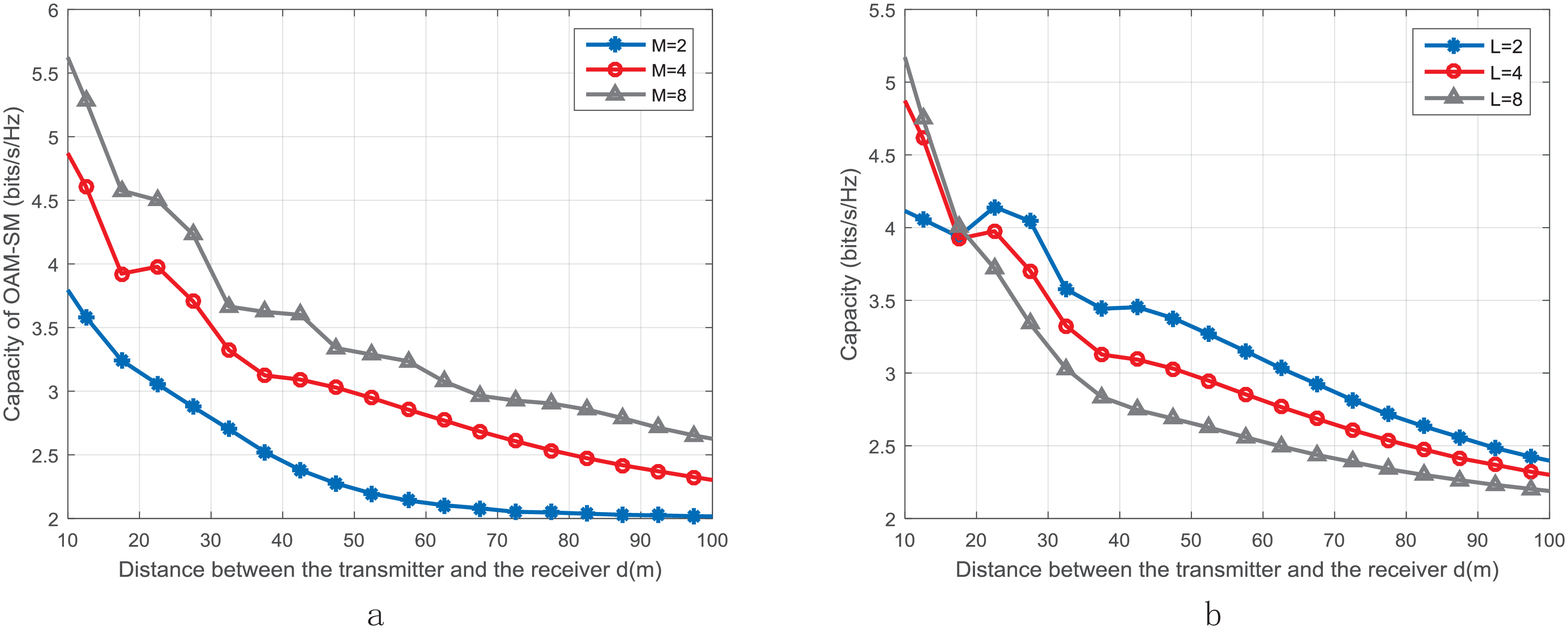}
\begin{quote}
\small Fig.6. Capacity of OAM-SM millimeter wave communication systems with respect to the transmission distance considering different numbers of transmission antennas and OAM states.
\end{quote}
\end{figure}

\begin{figure}
\centering
\includegraphics[width=10cm,draft=false]{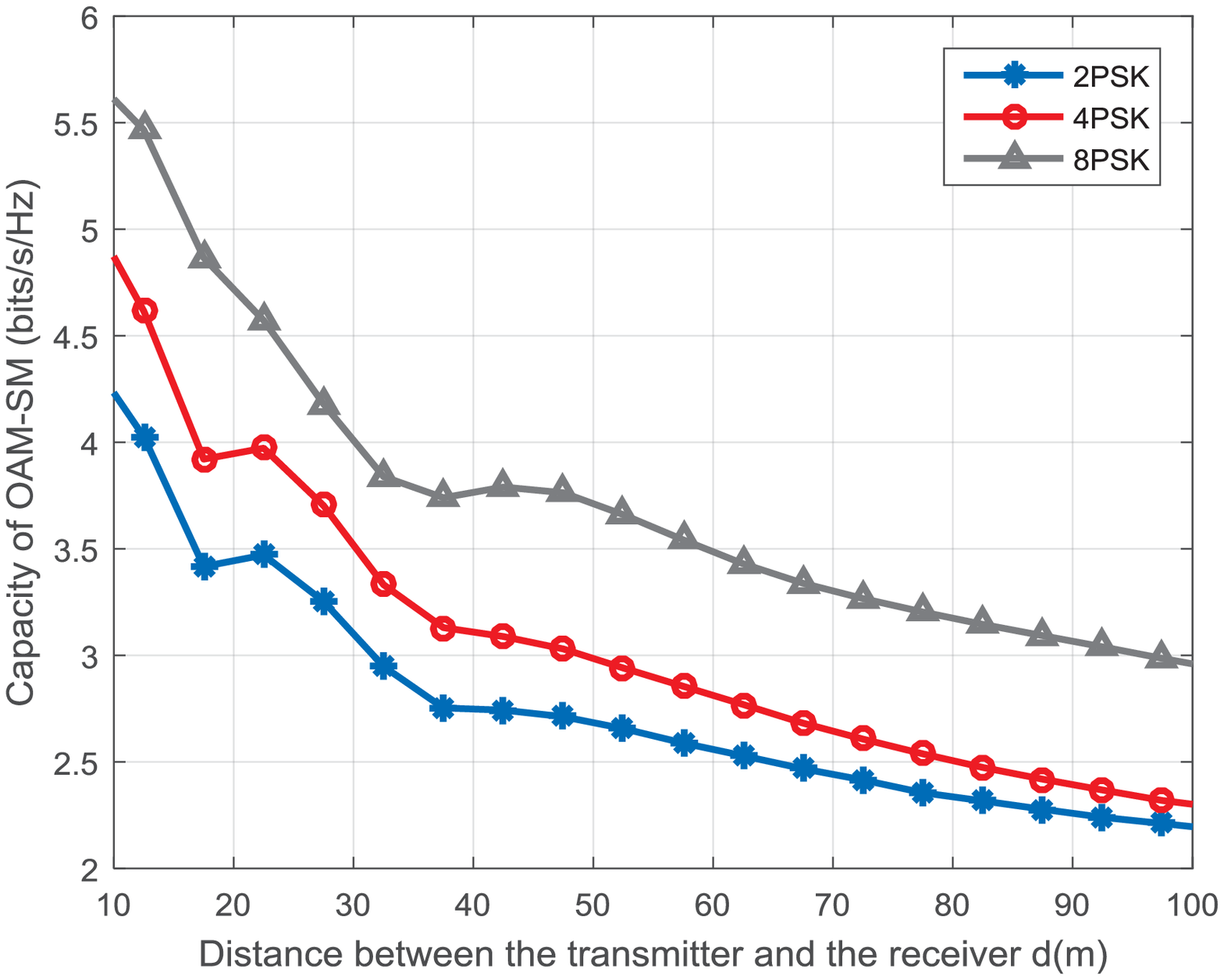}
\begin{quote}
\small Fig.7. Capacity of OAM-SM millimeter wave communication systems with respect to the transmission distance considering different $P$-point constellation modulation methods.
\end{quote}
\end{figure}

\begin{figure}
\centering
\includegraphics[width=10cm,draft=false]{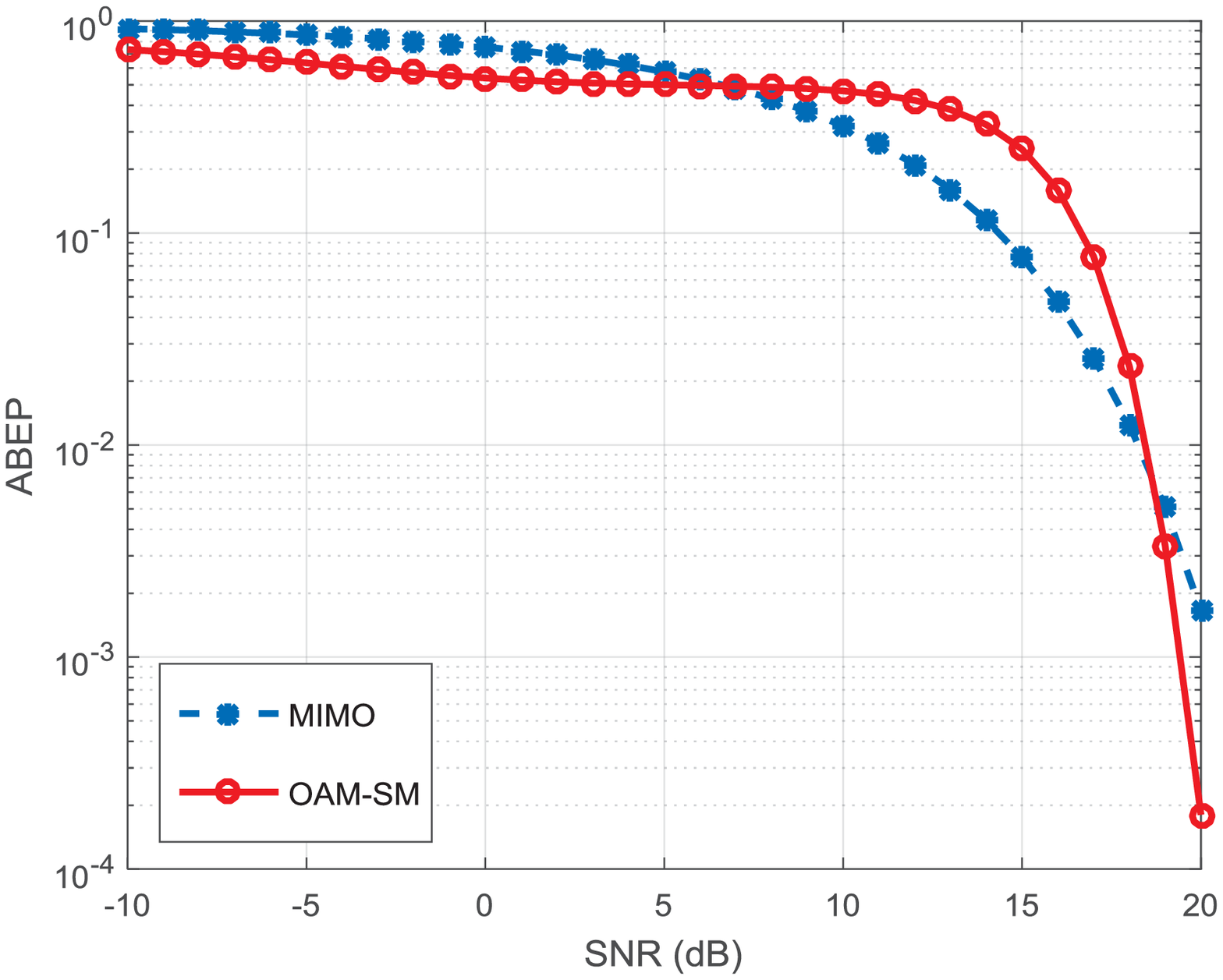}
\begin{quote}
\small Fig.8. ABEP of OAM-SM millimeter wave communication systems with respect to the transmission SNR.
\end{quote}
\end{figure}

\begin{figure}
\centering
\includegraphics[width=15cm,draft=false]{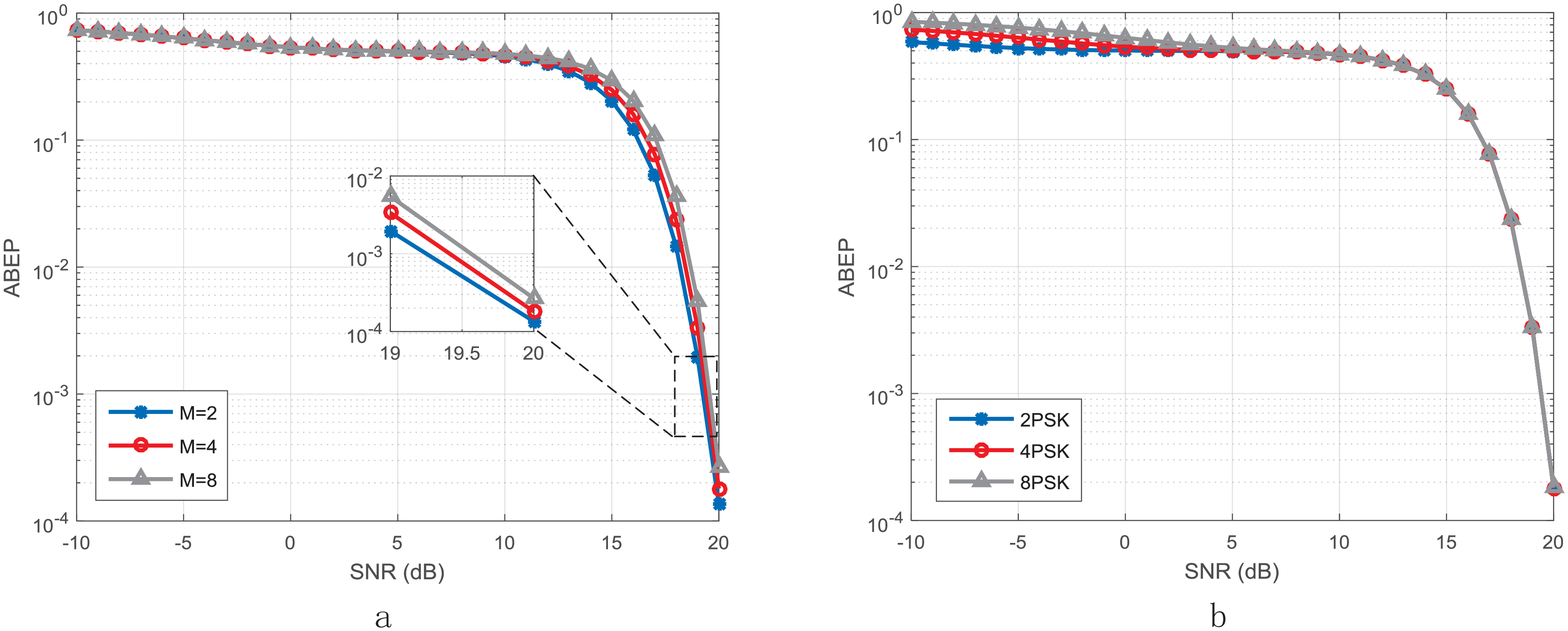}
\begin{quote}
\small Fig.9. ABEP of OAM-SM millimeter wave communication systems with respect to the transmission SNR considering different numbers of transmission antennas and $P$-point constellation modulation methods.
\end{quote}
\end{figure}

\begin{figure}
\centering
\includegraphics[width=15cm,draft=false]{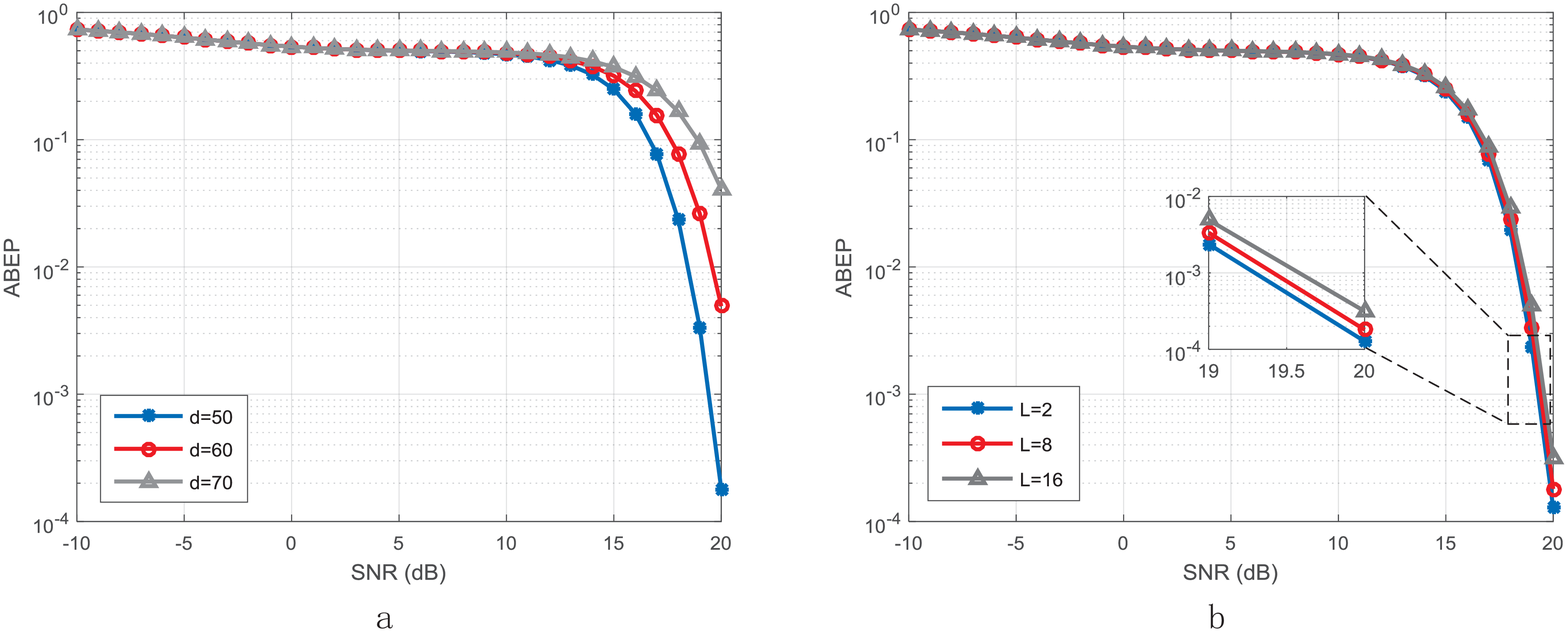}
\begin{quote}
\small Fig.10. ABEP of OAM-SM millimeter wave communication systems with respect to the transmission SNR considering different number of OAM states and distances between transmitters and receivers.
\end{quote}
\end{figure}

\begin{figure}
\centering
\includegraphics[width=10cm,draft=false]{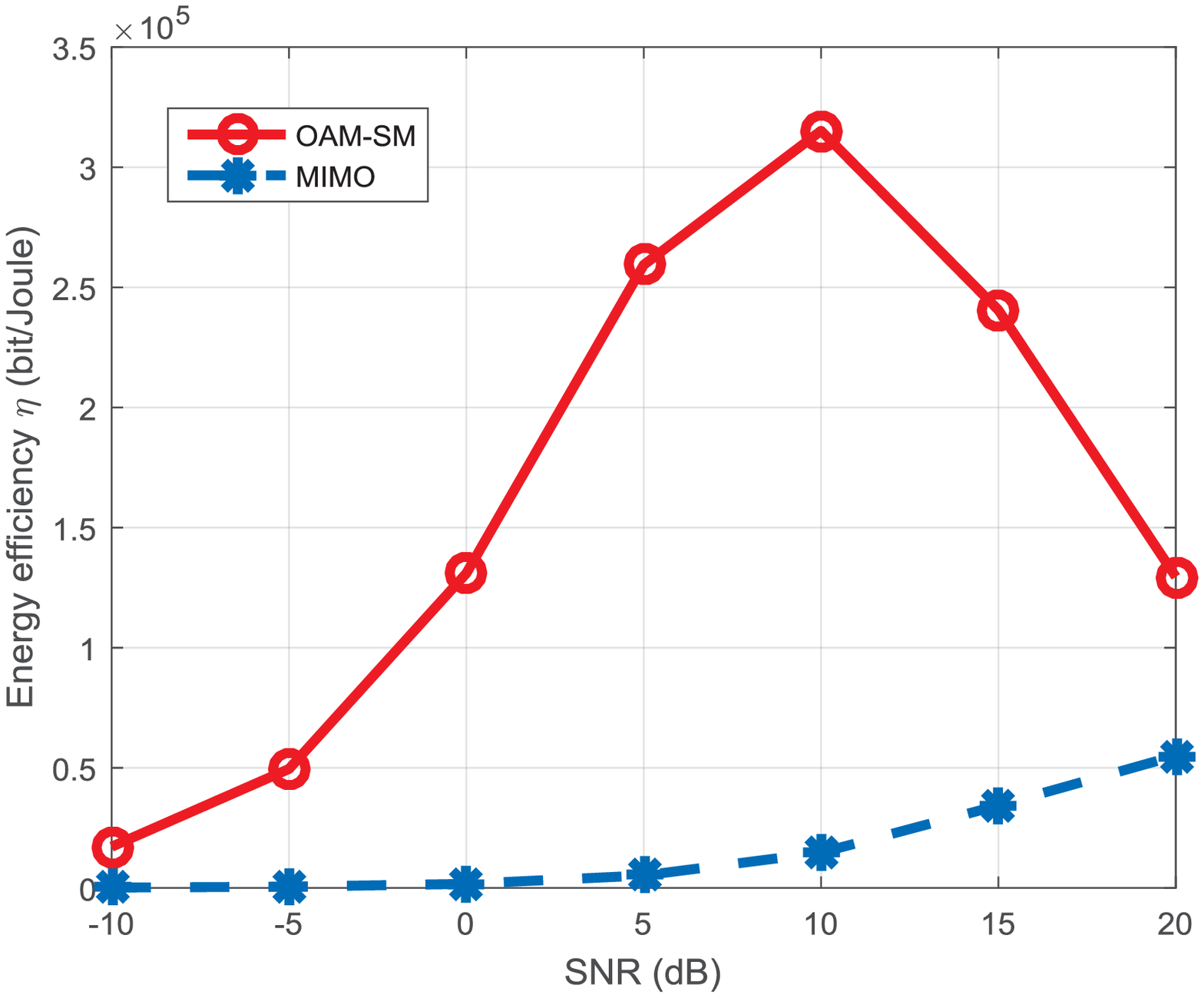}
\begin{quote}
\small Fig.11. Energy efficiency of OAM-SM and MIMO millimeter wave communication systems with respect to the transmission SNR.
\end{quote}
\end{figure}

\end{spacing}

\begin{thebibliography}{1}

\bibitem{Thompson14}
J. Thompson, X. Ge, H. C. Wu, et al. ``5G wireless communication systems: Prospects and challenges," in {\em IEEE Commun. Mag.}, vol. 52, No. 2, pp. 62--64, Feb. 2014.

\bibitem{Rappaport13}
T. S. Rappaport, S. Sun, R. Mayzus, et al. ``Millimeter wave mobile communications for 5G cellular: It will work!" in {\em IEEE Access}, Vol. 1, pp. 335--349, May. 2013.

\bibitem{Ge15}
X. Ge, S. Tu, T. Han, et al. ``Energy efficiency of small cell backhaul networks based on Gauss-Markov mobile models,"  in {\em IET Net.}, vol. 4, no. 2, pp. 158--167, Mar. 2015.

\bibitem{Akdeniz14}
M. R. Akdeniz, Y. Liu, M. K. Samimi, et al. ``Millimeter wave channel modeling and cellular capacity evaluation," in {\em IEEE J. Sel. Areas Commun.}, vol. 32, no. 16, pp. 1164--1179, Jun. 2014.

\bibitem{Yan14}
Y. Yan, G. X, P. J. Martin, et al. ``High-capacity millimetre-wave communications with orbital angular momentum multiplexing," in {\em Nature Commun.}, vol. 5, p. 4876, March. 2014.

\bibitem{Yao13}
A. Yao, M. J. Padgett. ``Orbital angular momentum: origins, behavior and applications," in {\em Adv. Opt. Photon.}, vol. 341, no. 6145, pp. 537--540, Aug. 2013.

\bibitem{Jackson98}
J. D. Jackson, {\em Classical Electrodynamics, 3rd ed.} New York: Wiley, 1998.

\bibitem{Gibson04}
G. Gibson, J. Courtial, M. P. Padgett, et al. ``Free-space information transfer using light beams carrying orbital angular momentum," in {\em Opt. Exp.}, vol. 12, no. 22, pp. 5448--5456, Nov. 2004.

\bibitem{Beth36}
R. A. Beth, ``Mechanical detection and measurement of the angular momentum of light," in {\em Phys. Rev.}, vol. 50, no. 2, pp. 115--125, Jul. 1936.

\bibitem{Allen92}
L. Allen , M. W. Beijersbergen, R. H. C. Spreeuw, et al. ``Orbital angular-momentum of light and the transformation of Laguerre-Gaussian laser modes," in {\em Phys. Rev.}, vol. 45, no. 11, pp. 8185--8189, Jun. 1992.

\bibitem{Awaji10}
Y. Awaji, N. Wada, Y. Toda, et al. ``Demonstration of spatial mode division multiplexing using Laguerre-Gaussian mode beam in telecom-wavelength," in {\em Proc. IEEE 23rd Annu. Meet. Photon. Soc.}, Denver. CO, Nov. 2010.

\bibitem{Djordjevic11}
I. B. Djordjevic, ``Deep-space and near-Earth optical communications by coded orbital angular momentum (OAM) modulation," in {\em Opt. Ecp.}, vol. 19, no. 15, pp. 14277--14289, Jul. 2011.

\bibitem{Wang11}
J. Wang, J. Yang, I. M. Fazal, N. Ahmed ``25.6-bit/s/Hz spectral efficiency using 16-QAM signals over pol-muxed multiple orbital-angular-momentum modes," in {\em Proc. IEEE Photon. Conf.}, Arlington, VA, USA, 2011, pp. 587--588.

\bibitem{Fazal11}
I. M. Fazal, J. Wang, J. Yang, et al. ``Demonstration of 2-Tbit/s data link using orthogonal orbital-angular-momentum modes and WDM," in {\em Proc. FiO/LS}, 2011, FTuT1.

\bibitem{Bozinovic13}
N. Bozinovic, Y. Yue, Y. Ren, et al. ``Terabit-scale orbital angular momentum mode division multiplexing in fibers," in {\em Science}, vol. 340, no. 6140, pp. 1545--1548, Jun. 2013.

\bibitem{Thid¨¦07}
B. Thid¨¦, H. Then, J. Sj\"{o}holm J, K. Palmer, et al. ``Utilization of photon orbital angular momentum in the low-frequency radio domain," in {\em Phys. Rev. Lett.}, vol. 99, no. 8, p. 087701, Aug. 2007.


\bibitem{Tamburini12}
F. Tamburini, E. Mari, A. Sponselli, et al. ``Encoding many channels in the same frequency through radio vorticity: first experimental test,"  in {\em New Journal of Physics}, vol. 14, no. 3, p. 033001, Mar. 2012.


\bibitem{Edfors12}
O. Edfors, A. J. Johansson, ``Is orbital angular momentum (OAM) based radio communication an unexploited area?," in {\em IEEE Trans. Ant. Pro.}, vol. 60, no. 2, PP. 1126--1131, Feb. 2012.

\bibitem{Zhang13}
Y. Zhang, W. Feng, N. Ge, ``On the restriction of utilizing orbital angular momentum in radio communications," in {\em Proc. VIII Int. Conf. on Communications and Networking in China (CHINACOM)}, Guilin, China, 2013.

\bibitem{Andersson15}
M. Andersson, E. Berglind, and G. Bj\"{o}rk, ``Orbital angular momentum modes do not increase the channel capacity in communication links," in {\em New.J. Phys.}, vol. 17, no. 4, p. 043040, Apr. 2015,

\bibitem{Hui15}
X. Hui, S. Zheng, Y. Chen, et al.  ``Multiplexed millimeter wave communication with dual orbital angular momentum (OAM) mode antennas," in {\em Sci. Rep.}, vol. 5, p. 10148, 2015.

\bibitem{Zhang16}
Z. Zhang, S. Zheng, Y. Chen, et al.  ``The capacity gain of orbital angular momentum based multiple-input-multiple-output system," in {\em Sci. Rep.}, vol. 6, p. 25418, May. 2016.

\bibitem{Tian16}
H. Tian, Z. Liu, W. Xi, et al. ``Beam axis detection and alignment for uniform circular array-based orbital angular momentum wireless communication," in {\em IET Commun.}, vol. 10, no. 1, pp. 44--49, Feb. 2016.

\bibitem{Allen14}
E. Allen, A. Tennant, Q. Bai, et al.  ``Wireless data encoding and decoding using OAM modes," in {\em Electronics Letters}, vol. 50, no. 3, pp. 232--233, Jan. 2014.

\bibitem{Oldoni15}
M. Oldoni, P. Spinello, E. Mari, et al. ``Space-division demultiplexing in orbital-angular-momentum-based MIMO radio systems," in {\em IEEE Trans. Ant. Pro.}, vol. 63, no. 10, pp. 4582--4587, Oct. 2015.

\bibitem{Mohammadi10}
S. M. Mohammadi, et al. ``Orbital angular momentum in radio: Measurement methods," in {\em Radio Sci.}, vol. 45, p. RS4007, Jul. 2010.

\bibitem{Gallager68}
R. Gallager, Information Theory and Reliable Communication. New York: John Wiley and Sons, 1968.

\bibitem{Fano61}
R. M. Fano, Transmission of Information: A statistical Theory of Communications. New York: John Wiley and Sons, 1961.

\bibitem{Yang08}
Y. Yang, B. Jiao. ``Information-guided channel-hopping for high data rate wireless communication," in {\em IEEE Commun. Lett.}, vol. 12, no. 4, pp. 1089--7789, Apr. 2008.

\bibitem{Goldsmith03}
A. Goldsmith, S.A. Jafar, N. Jindal, et al. ``Capacity limits of MIMO channels," in {\em IEEE J. Sel. Areas Commun.}, vol. 21, no. 5, pp. 684--702, Jun. 2003.

\bibitem{Mao13}
R. Mao, G. Mao, ``Road traffic density estimation in vehicular networks," in {\em Wireless Commun. and Netw. Conf.} (WCNC), pp. 4653--4658, Apr. 2013.

\bibitem{Mao06}
G. Mao, B. D. O. Anderson, and B. Fidan, ``WSN06-4: Online calibration of path loss exponent in wireless sensor networks," in {\em Proc. IEEE GLOBECOM}, pp. 1--6, Nov. 2006.

\bibitem{Xiang13}
L. Xiang, X. Ge,  C.-X. Wang, et al. `` Energy efficiency evaluation of cellular networks based on spatial distributions of traffic load and power consumption," in {\em IEEE Trans. Wire. Commun.}, vol. 12, no. 3, pp.  961--973, Mar. 2013.

\bibitem{Mood74}
Mood, A. Mcfarlane, et al. Introduction to the Theory of Statistics. 3rd ed. McGraw-Hill, 1974.

\bibitem{Zhang15}
R. Zhang, L.-L Yang, L. Hanzo, et al. ``Error probability and capacity analysis of generalised pre-coding aided spatial modulation," in {\em IEEE Trans. Wire. Commun.}, vol. 14, no. 1, pp. 364--375, Jan. 2015.

\bibitem{Irshid91}
M. I. Irshid, I.S. Salous. ``Bit error probability for coherent M-ary PSK system," in {\em IEEE Trans. Wire. Commun.}, vol. 39, no. 3, pp. 349--352, Mar. 1991.

\bibitem{Simon05}
M. K. Simon, M. S. Alouini. Digital communication over fading channels. 2nd ed. John Wiley and Sons, 2005.

\bibitem{Stavridis12}
A. Stavridis, S. Sinanovi, M. D. Renzo,  et al. ``Energy saving base station employing spatial modulation," in {\em Proc. IEEE 17th Int. Workshop CAMAD}, pp. 231--235, Sep. 2012.

\bibitem{Serafimovski13}
N. Serafimovski, et al. ``Practical implementation of spatial modulation," in {\em IEEE Trans. Veh. Technol.}, vol. 62, no. 9, pp. 4511--4523, Nov. 2013.

\bibitem{Auer11}
G. Auer, V. Giannini, I. Godor, P. Skillermark, et al. ``Cellular energy efficiency evaluation framework," in {\em Proc. 2011 IEEE Veh. Technol. Conf. -- Spring}, pp. 1--6,



\end{thebibliography}
\end{document}